\documentclass[11pt]{article}
\usepackage{times}
\usepackage{sidecap}
\usepackage{geometry}
\usepackage{floatrow}
\usepackage{float}
\usepackage{hyperref}
\usepackage{graphics,graphicx}
\usepackage{subcaption}
\usepackage{epsf,wrapfig}
\usepackage{natbib,bm}
\usepackage[misc]{ifsym}
\usepackage{bbding}

\usepackage{paralist}
\let\olditem\item
\renewenvironment{thebibliography}[1]{%
  \section*{\refname}
  \let\par\relax
  \renewcommand{\item}[1][]{\olditem[\textbullet]}%
  \inparaenum}{\endinparaenum}

\newcounter{aff}

\DeclareGraphicsExtensions{.jpg,.pdf,.png,.eps}

\usepackage{caption}
\usepackage[dvipsnames]{xcolor}
\usepackage{tcolorbox}
\bibpunct{(}{)}{;}{a}{}{,}
\usepackage{soul}
\usepackage[version=4]{mhchem}

\definecolor{darkcyan}{rgb}{0.0, 0.0, 0.7}

\definecolor{boxcolor}{RGB}{193, 229, 255}
\definecolor{figboxcolor}{RGB}{193, 229, 255}
\definecolor{boxcolorgray}{RGB}{227, 227, 227}
\definecolor{darkcyan}{rgb}{0.0, 0.0, 0.7}

\sethlcolor{boxcolorgray}

\def\gtrsim{\lower 2pt \hbox{$\, \buildrel {\scriptstyle >}\over
{\scriptstyle \sim}\,$}}
\def\lesssim{\lower 2pt \hbox{$\, \buildrel {\scriptstyle <}\over
{\scriptstyle \sim}\,$}}

\def\approxlt{\lower.2em\hbox{$\buildrel < \over \sim$}}
\def\approxgt{\lower.2em\hbox{$\buildrel > \over \sim$}}

\def\be{\begin{equation}}
\def\ee{\end{equation}}
\def\bea{\begin{eqnarray*}}
\def\eea{\end{eqnarray*}}

\def\jref@jnl#1{{\rm#1}}
\def\aj{\jref@jnl{AJ}}			
\def\araa{\jref@jnl{ARA\&A}}		
\def\apj{\jref@jnl{ApJ}}			
\def\apjl{\jref@jnl{ApJ}}		
\def\apjs{\jref@jnl{ApJS}}		
\def\ao{\jref@jnl{Appl.~Opt.}}		
\def\apss{\jref@jnl{Ap\&SS}}		
\def\aap{\jref@jnl{A\&A}}		
\def\aapr{\jref@jnl{A\&A~Rev.}}		
\def\aaps{\jref@jnl{A\&AS}}		
\def\azh{\jref@jnl{AZh}}			
\def\baas{\jref@jnl{BAAS}}		
\def\jrasc{\jref@jnl{JRASC}}		
\def\memras{\jref@jnl{MmRAS}}		
\def\mnras{\jref@jnl{MNRAS}}		
\def\pra{\jref@jnl{Phys.~Rev.~A}}	
\def\prb{\jref@jnl{Phys.~Rev.~B}}	
\def\prc{\jref@jnl{Phys.~Rev.~C}}	
\def\prd{\jref@jnl{Phys.~Rev.~D}}	
\def\pre{\jref@jnl{Phys.~Rev.~E}}	
\def\prl{\jref@jnl{Phys.~Rev.~Lett.}}	
\def\pasp{\jref@jnl{PASP}}		
\def\pasj{\jref@jnl{PASJ}}		
\def\qjras{\jref@jnl{QJRAS}}		
\def\skytel{\jref@jnl{S\&T}}		
\def\solphys{\jref@jnl{Sol.~Phys.}}	
\def\sovast{\jref@jnl{Soviet~Ast.}}	
\def\ssr{\jref@jnl{Space~Sci.~Rev.}}	
\def\zap{\jref@jnl{ZAp}}			
\def\nat{\jref@jnl{Nature}}		
\def\iaucirc{\jref@jnl{IAU~Circ.}}
\def\aplett{\jref@jnl{Astrophys.~Lett.}}
\def\apspr{\jref@jnl{Astrophys.~Space~Phys.~Res.}}
\def\bain{\jref@jnl{Bull.~Astron.~Inst.~Netherlands}}
\def\fcp{\jref@jnl{Fund.~Cosmic~Phys.}}
\def\gca{\jref@jnl{Geochim.~Cosmochim.~Acta}}
\def\grl{\jref@jnl{Geophys.~Res.~Lett.}}
\def\jcp{\jref@jnl{J.~Chem.~Phys.}}	
\def\jgr{\jref@jnl{J.~Geophys.~Res.}}	
\def\jqsrt{\jref@jnl{J.~Quant.~Spec.~Radiat.~Transf.}}
\def\memsai{\jref@jnl{Mem.~Soc.~Astron.~Italiana}}
\def\nphysa{\jref@jnl{Nucl.~Phys.~A}}
\def\physrep{\jref@jnl{Phys.~Rep.}}
\def\physscr{\jref@jnl{Phys.~Scr}}
\def\planss{\jref@jnl{Planet.~Space~Sci.}}
\def\procspie{\jref@jnl{Proc.~SPIE}}

\def\ion#1#2{#1$\;${\small\rm\@Roman{#2}}\relax}

\newcount\lecurrentfam
\def\LaTeX{\lecurrentfam=\the\fam \leavevmode L\raise.42ex
\hbox{$\fam\lecurrentfam\scriptstyle\kern-.3em A$}\kern-.15em\TeX}
\def\sizrpt{
(\fontname\the\font): em=\the\fontdimen6\font, ex=\the\fontdimen5\font
\typeout{
(\fontname\the\font): em=\the\fontdimen6\font, ex=\the\fontdimen5\font
}}
\definecolor{dark}{rgb}{ 0.34, 0.09, 0.49}

\usepackage{fancyhdr,url}
\usepackage{transparent}
\usepackage{tablefootnote}
\usepackage{mathpazo}

\begin{document}

\noindent
\begin{raggedright}
{\Huge\bfseries Chemistry on solar system scales \\[0.5em]
\Large \textit{What is the composition of terrestrial planet-building material?}
}\end{raggedright}\\\\
\noindent
\noindent
Catherine Walsh$^{1,}$\textsuperscript{\Letter}, 
Maria Drozdovskaya$^{2,}$\textsuperscript{\Letter}, 
Dmitry Semenov$^{3,}$\textsuperscript{\Letter}, 
Alice Booth$^{4,}$\textsuperscript{\Letter},
Margot Leem\-ker$^{5,}$\textsuperscript{\Letter},
Eleonora Bianchi$^{6}$,
Pooneh Nazari$^{7}$,
Linda Podio$^{6}$, 
Claudio Codella$^{6}$,
Giovanni Sabatini$^{6}$,
Elenia Pacetti$^{8}$,
Diego Turrini$^{9}$,
Danae Polychroni$^{9}$,
Myriam Benisty$^{10}$,
Claudia Toci$^{11}$,
Marta De Simone$^{12}$,
Víctor M.~Rivilla$^{13}$,
Enrique Macias$^{12}$,
Luna Rampinelli$^{5}$,
Beatrice Kulterer$^{14}$,
Claudio Hernández-Vera$^{15, 16}$,
Laura Colzi$^{13}$,
Hideko Nomura$^{17}$,
Maxime Tanious$^{18}$,
Ian Czekala$^{19}$,
Shota Notsu$^{20, 21}$,
Quentin Kral$^{22}$,
Charles Law$^{23}$,
Liton Majumdar$^{24,25}$,
Stefano Facchini$^{5}$,
Jenny Frediani$^{26}$,
Sebastiaan Krijt$^{27}$,
Teresa Paneque-Carreño$^{28}$,
\'Agnes K\'osp\'al$^{29}$,
S\'ebastien Maret$^{30}$, and 
Gianni Cataldi$^{17}$.

\vspace{1em}
{
\footnotesize{
\noindent $^1$ School of Physics and Astronomy, University of Leeds, UK; 
$^2$ Department of Chemistry, Biochemistry and Pharmaceutical Sciences, University of Bern, Switzerland; 
$^3$ ZAH/ITA, University of Heidelberg, Germany; 
$^4$ Centre for Astrophysics~\textbar~Harvard \& Smithsonian, USA; 
$^5$ Dipartimento di Fisica, Università degli Studi di Milano, Italy; 
$^6$ INAF/Arcetri Observatory, Italy; 
$^7$ Department of Physics, University of Warwick, UK; 
$^8$ INAF/IAPS, Italy; 
$^9$ INAF/OATO, Italy; 
$^{10}$ MPIA, Germany; 
$^{11}$ Departamento de Fisica Aplicada III, ETSI, Sevilla University, Spain; 
$^{12}$ European Southern Observatory, Garching, Germany; 
$^{13}$ Centro de Astrobiología (CAB), CSIC-INTA, Madrid, Spain; 
$^{14}$ Department of Astronomy, University of Virginia, USA; 
$^{15}$ European Southern Observatory, Santiago, Chile;
$^{16}$ YEMS, Chile;
$^{17}$ National Astronomical Observatory of Japan, Japan
$^{18}$ Institut de Ciències de l’Espai (ICE-CSIC), Spain; 
$^{19}$ Australia Telescope National Facility, CSIRO Space and Astronomy, Australia; 
$^{20}$ Department of Earth and Planetary Science, The University of Tokyo, Japan;
$^{21}$ Star and Planet Formation Laboratory, RIKEN, Japan;
$^{22}$ LIRA, Observatoire de Paris, Université PSL, Sorbonne Université, Université Paris Cité, CY Cergy Paris Université, CNRS, France; 
$^{23}$ Minnesota Institute for Astrophysics, University of Minnesota, USA;
$^{24}$ National Institute of Science Education and Research, Odisha, India;
$^{25}$ Homi Bhabha National Institute, Mumbai, India;
$^{26}$ Department of Astronomy, Stockholm University, Sweden;
$^{27}$ Department of Physics and Astronomy, University of Exeter, UK;
$^{28}$ Department of Astronomy, University of Michigan, USA;
$^{29}$ Konkoly Observatory, HUN-REN RCAES, Budapest, Hungary; 
$^{30}$ Univ.~Grenoble Alpes, CNRS, IPAG, France. \\

\noindent \Letter~\url{c.walsh1@leeds.ac.uk}; 
\Letter~\url{maria.drozdovskaya.space@gmail.com};
\Letter~\url{ dmitry.semenov@uni-heidelberg.de};\\
\Letter~\url{alice.booth@sao.si.edu};
\Letter~\url{margot.leemker@unimi.it}\\
}
}
\begin{tcolorbox}[boxsep=3pt,left=0pt,right=0pt,top=1pt,bottom=1pt,colframe=boxcolor,colback=boxcolor]
\justify
\textbf{Abstract:} The initial chemical composition of planets, primitive bodies in our solar system, and exoplanets is ultimately set by their nascent protoplanetary disks. Yet observing key carbon-, oxygen-, and nitrogen-bearing molecules (as well as rarer but important sulphur- and phosphorus-bearing species) in the inner (exo)planet-forming zone ($\lesssim 10$~au) remains extremely challenging with current facilities.
Therefore, to construct a complete picture of the chemistry of planet formation and understand how (exo)planetary composition is set by physical and chemical processes within the protoplanetary disk, a more powerful observing facility, i.e., ALMA2040, is needed. 
Compared to current ALMA (the Atacama Large Millimeter/submillimeter Array) this new facility must provide $\gtrsim$~a factor of 10 in line sensitivity at a spatial resolution of $\lesssim 5$~au at 140~pc, i.e., a factor of 3 - 4 better resolution to low surface brightness emission than currently possible. This distance includes many star-forming regions such as Taurus and Lupus that are host to populations of Sun-like stars.  
ALMA2040 will enable mapping of disk chemical composition on terrestrial planet-forming scales (within the water snowline) and detect molecules that are the feedstock for potentially habitable worlds.
\end{tcolorbox}

\newpage{}

\vspace{-0.7cm}
\section{Scientific context and open questions}
\label{introduction}
Planets form from dust, gas, and ice in the protoplanetary disks that surround newly-born stars \citep[e.g.,][]{Oberg2021b}. The chemistry and compositional gradients within these disks will determine the compositions of planetary systems, including potentially habitable worlds.
Key elements (and their heavier isotopes) for prebiotic chemistry
— C, H, N, O, P, and S —  are partitioned among different carriers, and their relative abundances will set the bulk atmospheric compositions of emerging planets \citep[e.g.,][]{Nomura2023}. 
In our solar system, the inner planets 
are small, rocky, and gas-poor, while the outer planets are large and gas-rich. This reflects the position of the snowline, beyond which water freezes out as ice, which facilitated the rapid growth of gas-giant cores. Other volatiles, e.g., \ce{CO2}, CO, and \ce{N2}, have unique snowlines that create prominent radial gradients in elemental ratios (e.g., C/O and N/O). 
These snowlines have left signatures in the atmospheres of Uranus and Neptune which are more enriched in heavy elements (C, O, and N) than those of Jupiter and Saturn \citep[e.g.,][]{Fortney10}.
Further, minor bodies, such as comets and asteroids, likely played an important role in shaping the composition of the terrestrial planets. Isotopic signatures preserved in Earth’s water and atmosphere show that impacts by volatile-rich minor bodies delivered material to the early Earth \citep[e.g.,][]{Altwegg2019}. Additionally, we can now make isotopic links between solar system bodies, interstellar objects, exoplanet atmospheres, and across the different stages and scales of star and planet formation \citep[e.g.,][]{Nomura2023,Huet2025, 2025ARA&A..63...83S,2026Natur.655..870C,Salazar2026,Salyk2026}. 

With over 6000 exoplanets now detected around other stars\footnote{\url{https://science.nasa.gov/exoplanets/discoveries-dashboard/}},
the time is ripe to characterize the composition of (exo)planet-forming zones. 
However, observing the composition of disks is challenging, because they are small objects ($\sim 100$~au or less in radius) at far distances ($\gtrsim 100$~pc) requiring high angular resolution observations to spatially resolve chemical gradients. 
Furthermore, most of the disk is cold ($\approx~20-100$~K) with ice dominating the volatile budget beyond the water snowline ($\approx$ a few au around a Sun-like star), requiring high sensitivity ($\lesssim 1$ mJy/beam) and a small beam size ($\approx 30$~mas) to detect gas-phase molecules and their rare isotopologues within the water snowline in star-forming regions that are host to Sun-like stars ($\approx$~140~pc).

The most powerful (sub)mm telescope currently available, ALMA (the Atacama Large Millimeter/submillimeter Array)\footnote{\url{https://www.almaobservatory.org/en/home/}}, has enabled mapping of rotational line emission of key elemental carriers (e.g., CO, HCN, and \ce{C2H}) at spatial scales on the order of 10 to 50~au, but only for the brightest \citep[e.g.,][]{Oberg2021a} and closest disks \citep[e.g., TW~Hya at 60~pc;][]{Nomura2021}, and in nearby star-forming regions \citep[e.g., Taurus-Aurigae at $140\pm20$~pc;][]{Garufi2021}.
To significantly advance our understanding of the composition and chemical diversity of planetary systems around other stars, 
we require a new (sub)mm facility to answer the following open questions:
\begin{tcolorbox}[boxsep=3pt,left=0pt,right=0pt,top=1pt,bottom=1pt,colframe=boxcolor,colback=boxcolor]
\begin{enumerate}
\item[A.] How diverse is the composition of planet-building material interior to the water snowline where terrestrial, and thus potentially habitable planets, form?
\item[B.] What are the isotopic ratios of key elements in the planet-forming regions around other stars, and how do they compare with the isotopic record of our solar system and other exoplanetary systems?
\item[C.] How chemically complex is planet-forming material, and do other planetary systems host the necessary ingredients for seeding habitability on terrestrial worlds? 
\end{enumerate}
\end{tcolorbox}

\section{Science case for ALMA2040}
\label{sciencecase}
\textcolor{RoyalBlue}{{\bf A.~Elemental abundances in the terrestrial planet-forming zones of solar nebula analogues}} - 
We are currently limited to mapping the composition of key elemental carriers only across the gas-giant planet-formation region, on scales of $\approx 15 - 20$ au at a distance of 140~pc \citep[e.g.,][]{Oberg2021a}. 
This is also only possible for exceptionally large or massive disks or those that have a favourable edge-on inclination that allows a direct view of the planet-forming midplane \citep[e.g., the `Flying Saucer';][]{Dutrey2025}. 
However, recent surveys show that the majority ($\approx 67\%$) of disks are small, spanning less than 30~au in radius in dust emission \citep[e.g.,][]{Guerra-Alvarado2025}.
These constraints significantly limit our understanding of planet formation in the most common environments and on the spatial scales at which rocky, and thus potentially habitable, planets form.
A new (sub)mm facility, such as ALMA2040, is needed to map the composition of key elemental carriers of carbon, oxygen, nitrogen, and sulphur, on terrestrial planet-building scales ($\lesssim 5$~au; $\lesssim$ 35~mas at 140~pc), and to increase the chances to detect phosphorus-bearing molecules, another key element for life. \\

\noindent \textcolor{RoyalBlue}{{\bf B.~Tracking the history of volatiles through their isotopic records}} - By detecting and mapping molecular line emission from rare isotopologues of key elemental carriers, current technology allows us to estimate the ratios of D/H, $^{13}$C/$^{12}$C, $^{18/17}$O/$^{16}$O, $^{15}$N/$^{14}$N, and $^{34/33}$S/$^{32}$S in a handful of nearby protoplanetary disks only~\citep[e.g.,][]{Nomura2023}.
These data are critical for determining the degree of chemical processing of the pristine pre-disk matter inherited from the parental molecular cloud \citep[e.g.,][]{Ceccarelli2014}, and for comparison with isotopic signatures present in meteorites, asteroids, and comets in our solar system that delivered material to the young Earth \citep[e.g.,][]{Alexander2018}, as well as with those in mature exoplanets \citep{2025ARA&A..63...83S}, and interstellar objects \citep{Salazar2026}.
Of particular importance is determining the water budget available for building potentially habitable `Earth-like' worlds around other solar-type stars, and measuring its isotopic ratios to determine whether the chemical origin and history of extrasolar water are similar to or different from those in our own solar system.
Rotational line emission from rare isotopologues is faint; hence, a new (sub)mm facility such as ALMA2040 with an order of magnitude increase in line brightness sensitivity is needed to measure isotopologue ratios in larger samples within nearby star-forming regions and trace the diversity of planet-forming environments beyond the handful of objects possible so far (see also Colzi et al.~ALMA2040).  \\

\begin{figure}
    \includegraphics[width=\textwidth]{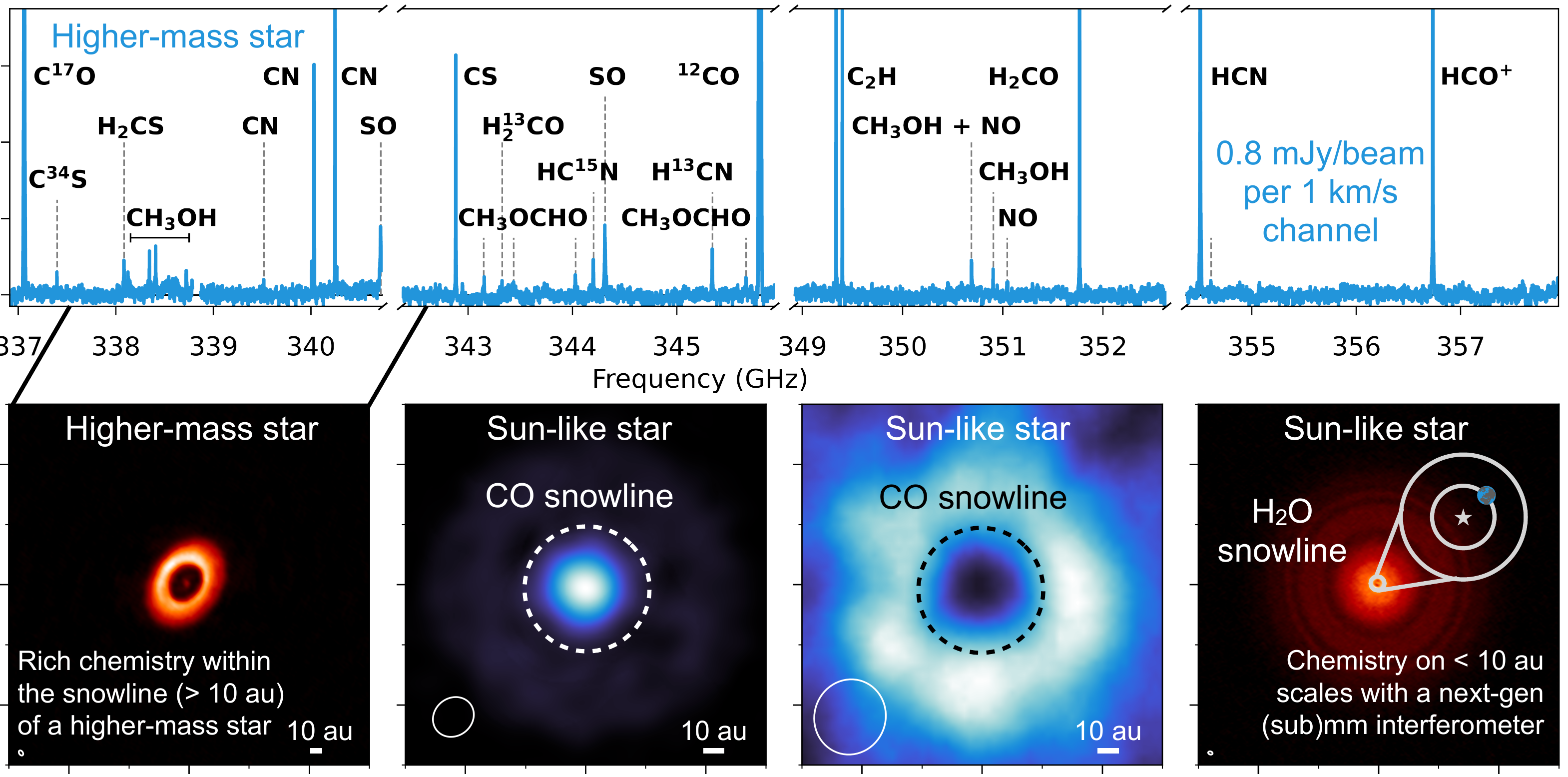}
    \caption{Current ALMA can detect the rich chemistry within the snowline around higher-mass stars, with an example spectrum shown in the top panel. For Sun-like stars, mapping across the CO snowline only ($\gtrsim 10$~au) is currently possible, which corresponds to the giant planet formation zone. The step changes in spatial resolution and sensitivity provided by ALMA2040 will allow us to detect and map the composition of terrestrial-planet-building material around Sun-like stars. Data from \citet{qi2013}, \citet{Andrews2016}, and \citet{Booth2024b}. \label{fig:current-alma}}
\end{figure}

\noindent \textcolor{RoyalBlue}{{\bf C.~Feedstock for pre-biotic chemistry during planet formation}} - 
Complex organic molecules (COMs) are species with six or more atoms and functional groups that are the building blocks of prebiotic chemistry (e.g., aldehydes (CHO), alcohols (OH), and carboxylic acids (COOH)). They have sublimation temperatures similar to or higher than, water \citep[e.g.,][]{Minissale2022}; hence, they are abundant in the gas phase within the water snowline only where their composition traces the ices delivered to the inner disk (see Fig.~\ref{fig:current-alma}). 
Hence, these COMs have only been detected in the inner regions of disks with sufficient masses of warm gas ($\gtrsim 100$~K), namely those around higher-mass stars \citep[$\approx 1.5 - 3~\mathrm{M}_\odot$; e.g.,][]{Booth2024a}, or young disks that have undergone a luminosity outburst \citep[][]{Lee2019}. These sources are not representative of the young solar system in which the snowline lay much closer to the star \citep[$\lesssim$~a few au; see, e.g., fig.~1 in][]{Lichtenberg2021}.
COMs are now routinely detected in the warm environments around forming stars (see also Rivilla et al.~ALMA2040), but it remains unknown how much of this material is inherited by planet-forming disks. This severely limits our understanding of whether or not the same ingredients that seeded life in our solar system via cometary delivery are also present around other stars \citep[e.g.,][]{Lippi2024}, and the degree of chemical processing that may occur as COMs are transported to and within the protoplanetary disk \citep[e.g.,][]{Drozdovskaya2016}.
Line emission from COMs is faint (\lesssim 10 mJy); hence detecting COMs in a large sample of disks with current facilities is not possible. A new (sub)mm facility such as ALMA2040 with an order of magnitude increase in sensitivity is needed to detect COMs in the disks around solar-type stars and constrain their abundances, heritage, and chemical diversity, by linking with protostellar and prestellar COM inventories (see also Drozdovskaya, Csengeri, et al.~ALMA2040).

\section{Case studies for ALMA2040}
\label{casestudies}

To show how ALMA2040 will provide the step change in sensitivity needed to meet our future science goals, we present synthetic spectra for two case studies around Sun-like stars: (1) detection of water within the snowline, and (2) detection of emission from COMs (carriers of key functional groups for prebiotic chemistry) in the planet-forming disk. We choose these case studies because detecting line emission on small spatial scales, and detecting weak molecular line emission, is currently achievable with ALMA for only a few (\lesssim 10) especially favourable objects; extending such detections to more typical disks around Sun-like stars would therefore represent a major leap forward for the field. The increase in sensitivity needed for both case studies will also enable mapping the emission from more abundant elemental carriers (e.g., CO, HCN, or \ce{C2H}) and their isotopologues within and across the terrestrial planet-forming zone. \\

\newpage
\noindent \textcolor{RoyalBlue}{{\bf Case study 1: determining the water budget within the water snowline around Sun-like stars}} - 
So far, water line emission has only been detected in disks with ALMA for a few sources: HL Tau \citep{Facchini2024}, V883~Ori \citep{Tobin2023, Leemker2025, Nakasone2026}, AS~205N \citep{Bosman2021}, and HD~100546 \citep{Rampinelli2026}. 
HL Tau is a young Class I source still in the act of forming and thus has a water snowline that lies at a significantly farther distance than expected in typical planet-forming disks \citep[$\lesssim~6$~au;][]{Facchini2024, Leemker2026}, only a single line is detected in AS~205N at low significance \citep{Carr2018}, V883~Ori is an outbursting disk which has a water snowline that has been pushed out to $\gtrsim~40$~au \citep{Cieza2016}, and HD~100546 is a Herbig-Ae disk which hosts a large dust cavity allowing direct irradiation, heating, and thus sublimation of ices at the inner cavity edge \citep[][]{Booth2021}.  
Whilst these detections have been important in advancing our understanding of water emission and its distribution, they are generally limited to objects that give us an advantageous view of their water emission. Extending these methods to larger, more representative samples of disks around Sun-like stars requires ALMA2040 with its planned step change in line-brightness sensitivity. 

\begin{figure}[t]
\includegraphics[width=\textwidth]{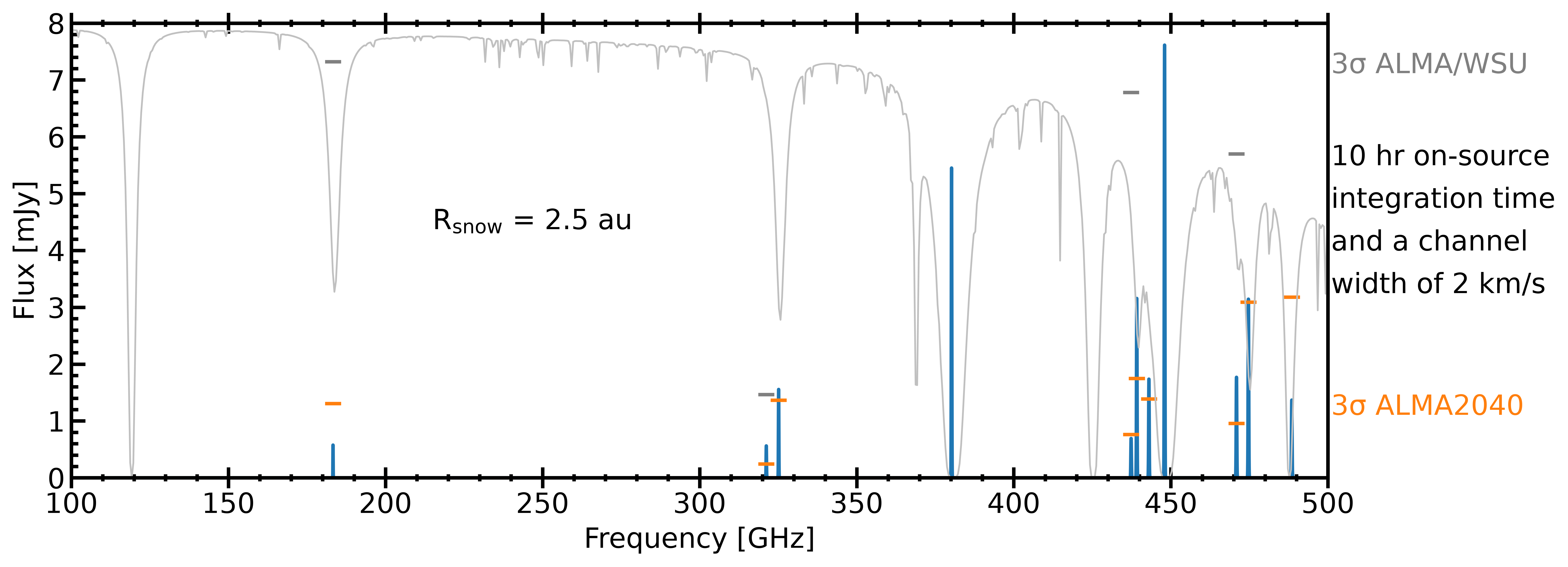}
\caption{Spectrum of \ce{H2O} for a protoplanetary disk around a Sun-like star at a distance of 140~pc (blue lines). 
The spectrum shows emission integrated within a single beam (35 mas) for which the radius corresponds to the location of the water snowline which we assume to lie at 2.5 au \citep[as proposed for the solar system;][]{Hayashi1981}. 
The grey dashes indicate the achievable sensitivity with current ALMA/WSU, while the orange dashes indicate what is achievable with ALMA2040 under optimal observing conditions. The grey curve indicates the transmission at the ALMA site assuming a pwv (precipitable water vapour) column of 0.472~mm.  
All water lines except for those at 183, 380, 448, and 488 GHz are predicted to be observable with ALMA2040 only.
\label{fig:dali}}
\end{figure}

To estimate the sensitivities needed to detect water within the snowline, 
we ran a thermochemical model of a protoplanetary disk around a Sun-like star using the DALI code, which allows computation of line emission in non-LTE \citep[Dust And LInes;][]{Bruderer2012,Bruderer2013}. 
We used ray-tracing assuming a face-on disk and molecular line data from LAMDA \citep[the Leiden Atomic and Molecular Database for Astrochemistry;][]{Schoier2005} to compute the line profiles for \ce{H2O}
integrated within the water snowline from 100 to 500 GHz, assuming a distance of 140~pc (similar to that for nearby star-forming regions).  
The model predicts a midplane snowline radius of $\approx 0.6$~au, so we rescale the emission to a larger emitting area assuming a water snowline at the same location as that proposed for the solar system \citep[2.5~au;][]{Hayashi1981}. 

\begin{SCfigure}
\includegraphics[width=0.7\textwidth]{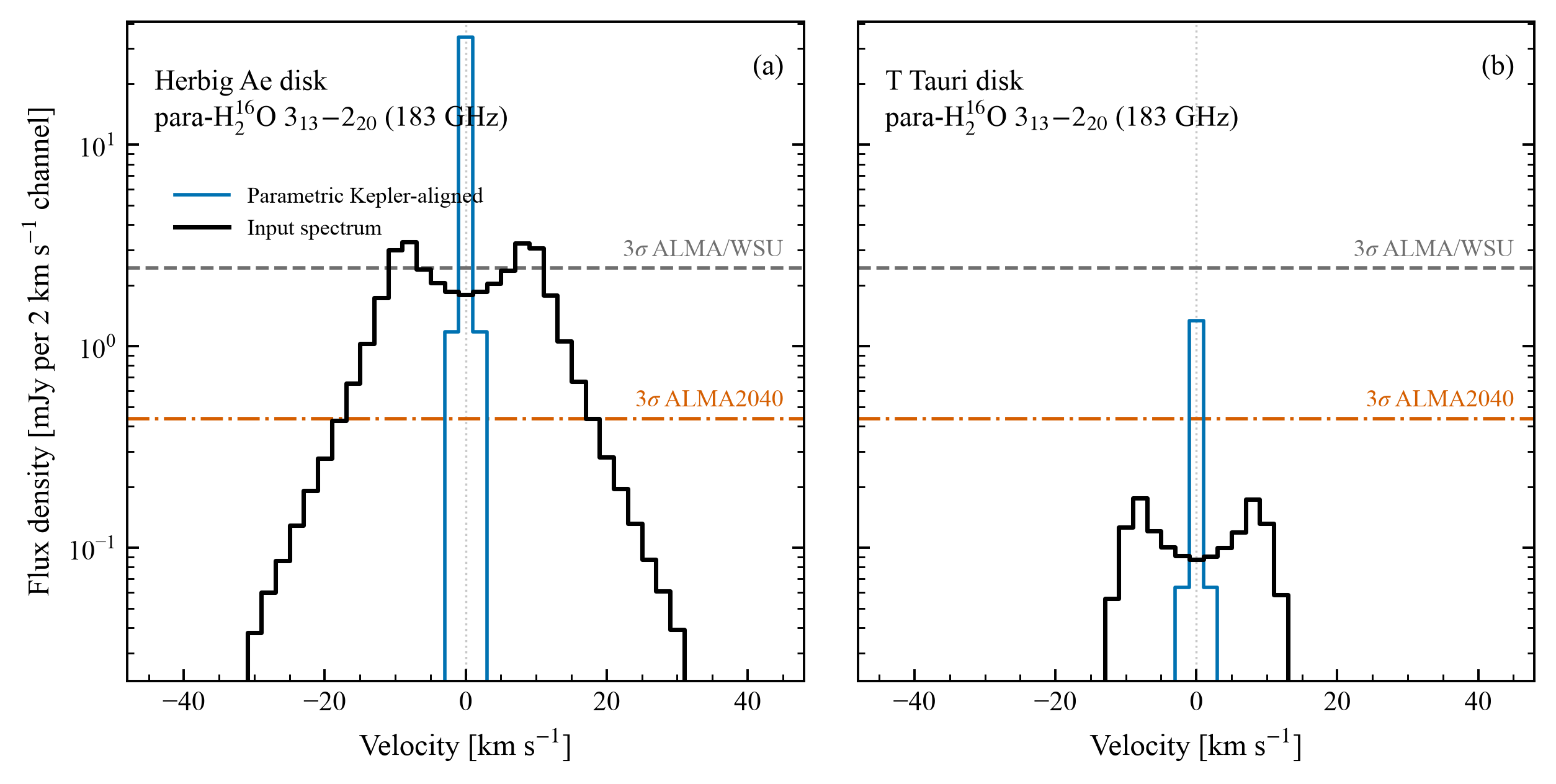}
\caption{Line profiles without (black) and with (blue) spectral stacking for a Herbig~Ae (left) and a T Tauri (right) disk for the 183 GHz water line. 
The expected 3$\sigma$ sensitivity for 10 hours of integration time and a channel width of 2 km/s is shown for ALMA/WSU (grey) and for ALMA2040 (orange). 
\label{fig:waterline}}
\end{SCfigure}

Figure \ref{fig:dali} shows the calculated water line spectrum within a single beam that is 35 mas in diameter.  
The grey curve shows the atmospheric transmission at the ALMA site for a 
precipitable water vapour (pwv) column of 0.472~mm. 
This corresponds to optimum observing conditions which are vital for remote sensing of water emission from astrophysical objects.  
The horizontal grey dashes show the ALMA/WSU (including the Wideband Sensitivity Upgrade) sensitivity with 10 hours of integration on source and a channel width of 2 km/s. 
Water lines within the snowline will be thermally broadened as well as rotationally broadened if the source is inclined; hence, this channel width gives us $> 10$ spectral resolution elements over the line (see Fig.~\ref{fig:waterline}).
The horizontal orange dashes, on the other hand, show the predicted ALMA2040 sensitivity for an equivalent observing time assuming an increase in array size to 255 12~m dishes.  
All water lines within this bandwidth will be observable with ALMA2040, except the lines at 183, 380, 448, and 488~GHz, while no lines are observable for the same observing time with current ALMA/WSU.   
{\textcolor{RoyalBlue}{\bf ALMA/WSU would need 15 days of integration to achieve the sensitivity required to detect the water line at 321~GHz with the same signal-to-noise ratio as 10 hours of integration with ALMA2040}} (see Fig~\ref{fig:dali}). 
The planned instantaneous coverage of $\approx 512$~GHz for ALMA2040 will also allow capturing of all water lines, spanning upper level energies from $\approx~200$ to $\approx 1,500$~K, in a single observation; hence, this adds an additional efficiency factor of a few above that achievable with ALMA/WSU. 
Overall \ce{H2O} in a disk around a solar-type star is observable only with ALMA2040, and the gains in both line-brightness sensitivity and bandwidth will enable the first population-level studies of water emission within the snowline in nearby star-forming regions.

The model predicts that the water line at 183~GHz remains unobservable even with ALMA2040; however, we can exploit signal-boosting techniques that use the known Keplerian velocity pattern of emission in inclined disks \citep[e.g.,][]{Yen2016}. 
Fig.~\ref{fig:waterline} shows the predicted line profile for the 183~GHz water line from a Herbig Ae (left) and a T Tauri (or solar-like; right) protoplanetary disk (black) and the predicted line flux when using spectral stacking \citep[blue;][]{Teague2019}. These data are from the high spectral resolution study of water line emission in disks by \citet{Notsu2018}. 
Without spectral stacking, this line remains undetected in a solar-like disk with current ALMA/WSU (grey line in Fig.~\ref{fig:dali}); however, with ALMA2040 and spectral stacking, this weak line now becomes well detected ($>3 \sigma$) in 10 hours of observing time (orange line). 
For systems in which the water snowline lies closer to the star (e.g., $\approx 0.6$~au as suggested by the DALI model), we can exploit signal-boosting techniques such as spectral stacking \citep[e.g.,][]{Teague2019} and line stacking \citep[e.g.,][]{Walsh2016} to increase the chances of detection. \\

\noindent {\textcolor{RoyalBlue}{\bf Case study 2: detecting complex organic molecules in solar analogue planet-forming disks}} - Whilst the detection of COMs is now routine around low-mass protostars, this remains challenging for more evolved planet-forming disks. This is because disks around Sun-like stars are colder, with significant freezeout of material beyond the water snowline, leading to lower gas-phase column densities and weaker line emission. To date, the only disk around a Sun-like star in which both O- and N-bearing COMs have been detected with ALMA is TW Hya, with successful detections of \ce{CH3OH} and \ce{CH3CN} \citep{Walsh2016,Loomis2018}, as well as the simplest organic acid, HCOOH \citep{Favre2018}. 
These detections have been facilitated by the proximity of TW~Hya \citep[60~pc;][]{Gaia2023}. 
Modelling suggests that the COM emission is arising from a luke-warm molecular layer driven by non-thermal desorption from ices \citep{Parfenov2017,Loomis2018,Calahan2023,Ilee2026}.
Non-thermal desorption (driven by e.g., stellar photons) does not give a direct view of the composition of ices, and this is only possible via {\em direct} sublimation within the snowline. 
The detection of COMs due to the direct sublimation of ices is limited to sources that offer us an advantageous view, including the outbursting disk of V883~Ori \citep[e.g.,][]{Lee2019, Yamato2024, Jeong2025} and the inner irradiated cavities of higher-mass Herbig Ae/Be disks \citep[e.g.,][and see Fig.~\ref{fig:current-alma}]{Booth2021,Booth2026}. 
These observations provide compelling evidence that the inner regions of disks around Sun-like stars may be similarly rich in COMs, but this remains untested, and abundance patterns could differ due to the impact of different host stars and disk evolution timescales.

To estimate the line sensitivity needed to detect molecules 
of potential prebiotic importance in disks around Sun-like stars, we present results from \citet{Walsh2014,Walsh2018} who performed ray-tracing calculations for key COMs with relevant functional groups (e.g., \ce{-NH2}, \ce{-COOH}, \ce{-OCH3}, and \ce{-HCO}).  
The molecular data used in the ray-tracing calculations were from CDMS \citep[the Cologne Database for Molecular Spectroscopy;][]{Muller2001}.
They used a template physical structure model for a disk around a T Tauri star that is 300~au in radius \citep{Nomura2007}, which was coupled with a gas-grain chemical model that included the formation and sublimation of COMs and computed the abundance of COMs as a function of disk radius and height \citep{Walsh2014}. Computed disk-integrated spectra were produced assuming a distance of 140~pc, which is representative of nearby star-forming regions. COM emission from disks is typically weak ($\lesssim 10$~mJy); hence disk-integrated spectra are calculated to maximise the predicted signal.

\begin{figure}[t]
\includegraphics[width=\textwidth]{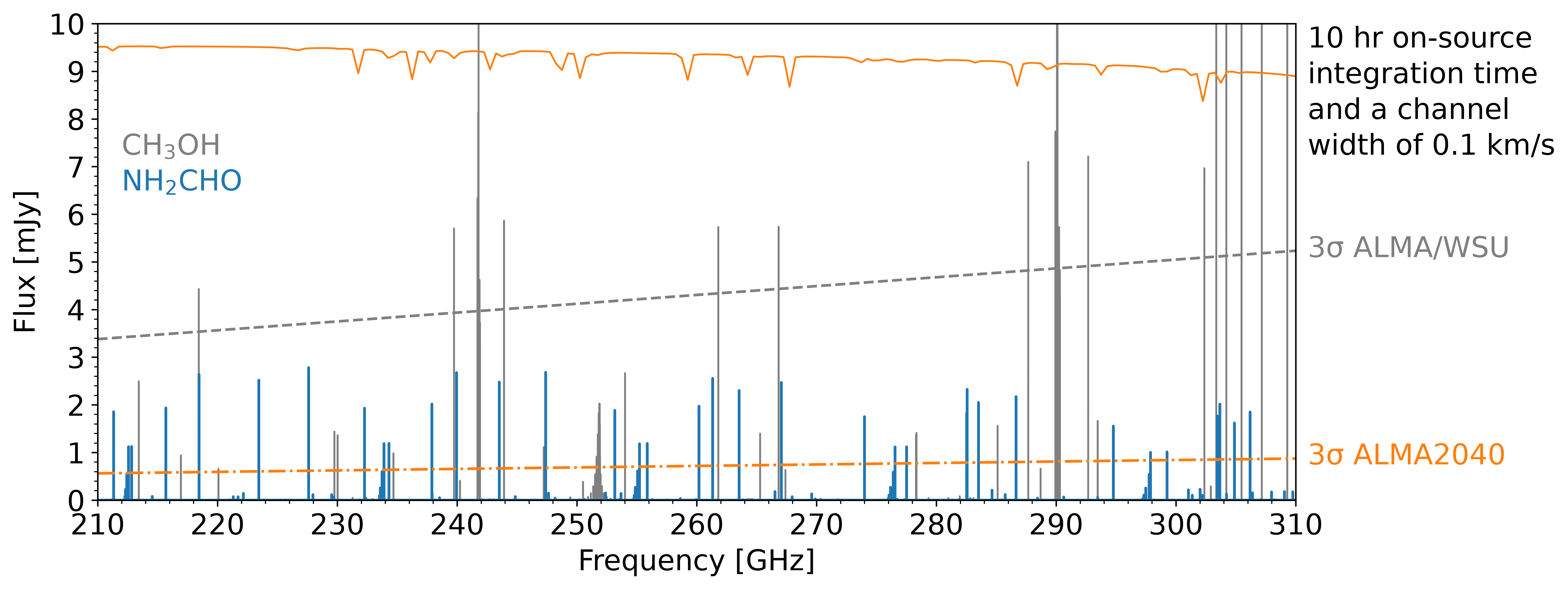}
\caption{Disk-integrated line spectra for \ce{CH3OH} (grey) and \ce{NH2CHO} (blue) for a disk around a T Tauri star \citep[from][]{Walsh2014,Walsh2018} from 210 to 310 GHz. 
The orange curve shows the atmospheric transmission at the ALMA site for typical observing conditions (0.913~mm pwv). 
The dashed lines show the expected sensitivity for current ALMA (grey) versus that for ALMA2040 (orange). Only ALMA2040 can reach the sensitivity needed to detect molecules of potential prebiotic importance, such as \ce{NH2CHO}. \label{fig:coms}}
\end{figure}

Figure~\ref{fig:coms} shows this synthetic spectrum over a wide band from 210--310 GHz which is a sweet spot for rotational line emission from COMs at moderate excitation temperatures ($\approx 20-100$~K). 
We show disk-integrated spectra for two illustrative species: \ce{CH3OH} (grey spectrum), which remains the only O-bearing COM detected in a disk around a solar-type star \citep{Walsh2016,Ilee2026}, and \ce{NH2CHO} (blue spectrum) which is the simplest molecule with a peptide bond, also present in amino acids, and thus considered an important precursor of prebiotic chemistry \citep[e.g.,][]{Jones2011}. So far an upper limit only for \ce{NH2CHO} is reported for the protoplanetary disk around HD~163296 \citep[][]{Kashyap2026}.
The grey horizontal dashed line indicates the expected 3$\sigma$ level for 10 hours of integration on source with ALMA/WSU, a channel width of 0.1 km~s$^{-1}$, and a precipitable water vapour column density of 0.913~mm (corresponding to typical observing conditions).  
The outer regions of protoplanetary disks are cool with narrow line widths ($\sim 1$~km/s) so we adopt a channel width that ensures at least 10 spectral channels across the line. 
The line slope is a guide for the eye to show the variation in expected sensitivity due to atmospheric transmission across this frequency range (shown by the orange curve). 
Only the strongest \ce{CH3OH} lines reach this detectability threshold, and none of the \ce{NH2CHO} lines are detectable with current ALMA/WSU.  
On the other hand, the orange dot-dashed line indicates the equivalent 3$\sigma$ level for ALMA2040 assuming an increase in array size to 255 12~m dishes.  
A forest of lines from \ce{NH2CHO} now appears above this detection threshold, meaning that, as well as a successful detection, the column density and excitation temperature of this species and other important COMs can be determined using a rotational diagram analysis \citep[as recently done for \ce{CH3OH};][]{Ilee2026}.   
The disk-integrated spectra in \citet{Walsh2018} suggest that emission from larger species such as \ce{CH3OCH3} and \ce{HOCH2CHO} should be detectable with ALMA2040 with peak line emission ranging from $\approx 1 - 3$ mJy. Line confusion (where multiple species can contribute to emission within a single channel) is not expected to be an issue for disks observed with ALMA2040 based on existing observations of COMs from Herbig disks with current ALMA \citep[][]{Booth2026}.

The planned instantaneous bandwidth of ALMA2040 ($\approx 512$~GHz) will capture an entire spectrum in one observation, increasing the telescope efficiency severalfold over current ALMA/WSU. If COM emission is weaker than predicted, signal-boosting techniques (e.g., line stacking) can be exploited to increase the chances of detection.
While the WSU will significantly increase the bandwidth of current ALMA and cover many line transitions in one observational setup, it will not provide the order-of-magnitude sensitivity {\em per channel} needed to detect weak molecular line emission; this will be possible only with ALMA2040.

\section{Synergy with other facilities}
\label{synergy}
ALMA2040 would be highly complementary to existing (e.g., JWST) and planned future facilities at other wavelengths (e.g., ELT, SKAO, ngVLA, and PRIMA).
JWST is providing unique insight into the hot water reservoir ($\gtrsim$~300~K; \citealt{2023Natur.620..516P}), in the surface layers of the inner regions of disks \citep[$\lesssim$~a~few~au; e.g.,][]{Banzatti2025}, and it can detect small infrared-active species \citep[e.g., \ce{CH4} and \ce{C2H2};][]{Tabone2023} that provides composition information inaccessible at (sub)mm wavelengths. 
ALMA2040, on the other hand, would enable detection of water in the deeper disk layers. 
The ELT will have numerous instruments from optical to near-infrared to observe emission from atomic and ionic lines tracing accretion and dispersal processes 
on spatial scales of $\sim 1$~au at a distance of 140~pc. 
PRIMA (PRobe far-Infrared Mission for Astrophysics) is a next-generation far-infrared space-borne telescope covering wavelengths from 24 to 235 $\mu$m (1.2 to 12.5~THz) with both spectroscopic and imaging modes.  
PRIMA will cover the fundamental rotational transitions of water and other small hydrides (e.g., HD) at frequencies only observable from space, providing a highly complementary frequency range to what is possible from the ground. 
Next-generation radio telescopes (e.g., SKAO and ngVLA) will operate at the lower frequencies required to detect heavy carbon-rich species \citep[e.g., \ce{C2H5CN};][]{Podio2026} and \ce{NH3}; however, (sub-)mm wavelengths remain optimal for detecting and mapping key elemental carriers in gas at temperatures $\gtrsim 20-50$~K. 
A new (sub)mm facility, such as ALMA2040, that approaches the spatial resolution of the ELT (down to a few mas) will provide observations of molecular emission from deeper and denser layers than accessible at optical and near-infrared wavelengths, at the scales at which disks also accrete and eject material.

\section{Technical requirements}
\label{technical}

To meet our science goals, ALMA2040 will require:
    \begin{itemize}
    \itemsep -0.3em
    \item a \textcolor{RoyalBlue}{{\bf spectral resolution of $0.1 - 1$~km/s}} to well resolve the narrow lines expected from protoplanetary disks; 
    \item a \textcolor{RoyalBlue}{{\bf spectral range from  35 to 500~GHz}} to ensure full coverage of molecular carriers of key elements including \ce{H2O};
    \item a \textcolor{RoyalBlue}{{\bf factor of  $\gtrsim10$ increase in line brightness sensitivity per channel}} to ensure the detection of weak lines from water, isotopologues, and COMs, and to enable mapping of strong emission from key elemental carriers across the terrestrial planet-forming zone;
    \item an \textcolor{RoyalBlue}{{\bf instantaneous bandwidth of $\approx 512$~GHz}} to capture a full disk spectrum in one observation and enabling excitation and
    abundance measurements, and;
    \item a \textcolor{RoyalBlue}{{\bf spatial resolution of $\approx 35$~mas}} at a distance of $\approx~140$~pc allowing at least two beams across the terrestrial planet-forming region and to resolve emission within and across the snowline.
    \end{itemize}

\noindent This new facility would also enable detections of rare isotopologues and COMs in smaller disks, and allow us to extend surveys to more distant star-forming regions and to disks beyond the local solar neighbourhood. 
This includes comprehensive surveys of key elemental carriers of carbon, oxygen, and nitrogen in regions like Orion (400~pc), where the presence of massive stars impacts disk evolution — much like the environment in which the Sun formed \citep[e.g.,][]{Adams2010}.

\begin{tcolorbox}[boxsep=3pt,left=0pt,right=0pt,top=1pt,bottom=1pt,colframe=boxcolor,colback=boxcolor]
\justify
\textbf{Summary:} A future facility, i.e., ALMA2040, with an order-of-magnitude increase in line sensitivity and an angular resolution matched to the terrestrial planet-forming zone, would transform our understanding of planet-formation chemistry. This facility would move the field from single-object case studies to statistically meaningful populations, from the handful of nearby `unicorn' disks currently accessible with ALMA to the full diversity of disks across star-forming regions out to 400~pc; hence, including the disk population shaped by external irradiation in Orion. For the first time, this would enable population-level comparisons of water budgets across the snowline, elemental and isotopic abundances of the key carbon-, nitrogen-, oxygen-, and sulphur- carriers, and complex organic molecule inventories across pre-stellar cores, protostars, planet-forming disks, and solar system and interstellar comets. This will allow us to directly link the chemical starting conditions of planet formation to its outcomes.
\end{tcolorbox}
\newpage

\small
\noindent\textsc{\textbf{Acknowledgements}}: {This work was developed as part of the broader ALMA2040 community effort. We thank an anonymous referee for their insightful and constructive comments. 
We acknowledge the Lorentz Center and NOVA for supporting the community workshops and coordination activities that helped initiate and shape the ALMA2040 White Paper process.
C.W.~acknowledges support from STFC and UKRI (grant numbers ST/X001016/1 and MR/Z00019X/1). 
D.S. was funded by DFG (German Research Foundation) – project number: 550639632. 
MND’s work is supported by ERC grant PSII (DOI:~10.3030/101\-230593).
A.S.B. gratefully acknowledges the support by the Clay Fellowship of the Smithsonian Astrophysical Observatory, made possible by the generosity of the Clay Family.
M.L.~is funded by the European Union (ERC, UNVEIL, 101076613).
E.P., D.P.~and D.T.~acknowledge support of the Italian Space Agency (ASI) through the ASI-INAF agreement no.~2021-5-HH.0 and addenda 2021-5-HH.1-2022 and 2021-5-HH.2-2024. 
E.P.~acknowledges support of the European Research Council via the Horizon 2020 Framework Programme ERC Synergy “ECOGAL” Project GA-855130, the ASI-INAF agreement no. 2024-5-HH.0, and the ASI-INAF agreement no. 2016-23-H.0 with its addendum 2016-23-H.3-2023. 
S.N.~is grateful for support from MEXT/JSPS Grants-in-Aid for Scientific Research (KAKENHI) grant Nos. JP23K13155, JP24K00674, and JP23H05441, and the Start-up Research Grant (University of Tokyo Excellent Young Researchers 2024). 
C.H.-V.~acknowledges support from the ANID -- Millennium Science Initiative Program -- Center Code NCN2024\_001. 
L.M.~acknowledges support from the DAE through the NISER project RNI 4011. 
A.K.~acknowledges support by the NKFIH NKKP grant ADVANCED 149943.
M.T.~acknowledges funding from the European Research Council (ERC) under the European Union’s Horizon 2020 research and innovation programme (Grant agreement No. 101098309 - PEBBLES). M.B.~has received funding from the European Research Council (ERC) under the European Union’s Horizon 2020 research and innovation programme (PROTOPLANETS, grant agreement No. 101002188). 
M.N.D.~, M.L.~, E.P., and M.P's work is funded by the European Union. Views and opinions expressed are however those of the author(s) only and do not necessarily reflect those of the European Union or the European Research Council Executive Agency. Neither the European Union nor the granting authority can be held responsible for them.
}
\\

\noindent
\bibliographystyle{apj_mod}
{\bibliography{chem_refs}}

@ARTICLE{Salazar2026,
       author = {{Salazar Manzano}, Luis E. and {Paneque-Carre{\~n}o}, Teresa and {Cordiner}, Martin A. and {Bergin}, Edwin A. and {Lin}, Hsing Wen and {Lis}, Dariusz C. and {Gerdes}, David W. and {Bergner}, Jennifer B. and {Biver}, Nicolas and {Bockel{\'e}e-Morvan}, Dominique and {Bodewits}, Dennis and {Charnley}, Steven B. and {Crovisier}, Jacques and {Farnocchia}, Davide and {Guzm{\'a}n}, Viviana V. and {Milam}, Stefanie N. and {Noonan}, John W. and {Remijan}, Anthony J. and {Roth}, Nathan X. and {Tobin}, John J.},
        title = "{Water D/H in 3I/ATLAS as a probe of formation conditions in another planetary system}",
      journal = {Nat. Astron.},
         year = 2026,
        month = jul,
       volume = {10},
        pages = {988-1000},
          doi = {10.1038/s41550-026-02850-5},
archivePrefix = {arXiv},
       eprint = {2603.07026},
 primaryClass = {astro-ph.EP},
       adsurl = {https://ui.adsabs.harvard.edu/abs/2026NatAs..10..988S}
}

@ARTICLE{Adams2010,
       author = {{Adams}, Fred C.},
        title = "{The Birth Environment of the Solar System}",
      journal = {\araa},
         year = 2010,
        month = sep,
       volume = {48},
        pages = {47-85},
          doi = {10.1146/annurev-astro-081309-130830},
archivePrefix = {arXiv},
       eprint = {1001.5444},
 primaryClass = {astro-ph.SR},
       adsurl = {https://ui.adsabs.harvard.edu/abs/2010ARA&A..48...47A}
}

@INCOLLECTION{Podio2026,
       author = {{Podio}, Linda and {Giani}, Lisa and {Walsh}, Catherine and {Coutens}, Audrey and {Jim{\'e}nez-Serra}, Izaskun and {Codella}, Claudio and {Maureira}, Mar{\'\i}a Jos{\'e} and {De Simone}, Marta and {Ilee}, John D. and {Lippi}, Manuela and {Lee}, Chin-Fei and {Le Gal}, Romane and {Narang}, Mayank and {Sabatini}, Giovanni and {Bianchi}, Eleonora and {Pacetti}, Elenia and {Polychroni}, Danai and {Banerjee}, Bihan and {Caselli}, Paola and {Ceccarelli}, Cecilia and {Farhang}, Amin and {Garufi}, Antonio and {Guidi}, Greta and {Ingallinera}, Adriano and {Ivanovski}, Stavro L. and {Klaassen}, Pamela and {L{\'o}pez-Sepulcre}, Ana and {Majumdar}, Liton and {Perotti}, Giulia and {Pineda}, Jaime E. and {Price}, Daniel J. and {Puravankara}, Manoj and {Rivi{\`e}re-Marichalar}, Pablo and {S{\'a}nchez-Monge}, Alvaro and {Schisano}, Eugenio and {Simonetti}, Paolo M. and {Testi}, Leonardo and {Toci}, Claudia and {Turrini}, Diego and {Traficante}, Alessio and {Wu}, Yinhao},
        title = "{Unveiling Complex Chemistry in Planet-forming Disks with the SKAO}",
    booktitle = {Advancing Astrophysics with the SKA I},
         year = 2026,
       editor = {{Bonaldi}, Anna and et al.},
    publisher = {SKAO},
          eid = {37},
        pages = {37},
          doi = {10.48550/arXiv.2606.27289},
       adsurl = {https://ui.adsabs.harvard.edu/abs/2026aask.bookE..37P}
}

@ARTICLE{Kashyap2026,
       author = {{Kashyap}, Parashmoni and {Majumdar}, Liton and {Bergin}, Edwin A. and {Blake}, Geoffrey A. and {Willacy}, Karen and {Guilloteau}, St{\'e}phane and {Dutrey}, Anne and {Liu}, Sheng-Yuan and {Henning}, Thomas and {Goldsmith}, Paul F. and {Lis}, Dariusz C. and {Maitrey}, S. and {Turner}, Neal and {Sahai}, Raghvendra and {Lee}, Chin-Fei and {Saito}, Masao},
        title = "{Unveiling the Chemical Complexity and C/O Ratio of the HD 163296 Protoplanetary Disk: Constraints from Multiline ALMA Observations of Organics, Nitriles, and Sulfur-bearing and Deuterated Molecules}",
      journal = {\apjs},
         year = 2026,
        month = feb,
       volume = {282},
       number = {2},
          eid = {47},
        pages = {47},
          doi = {10.3847/1538-4365/ae278e},
archivePrefix = {arXiv},
       eprint = {2511.10882},
 primaryClass = {astro-ph.EP},
       adsurl = {https://ui.adsabs.harvard.edu/abs/2026ApJS..282...47K}
}

@ARTICLE{Gaia2023,
       author = {{Gaia Collaboration} and {Vallenari}, A. and {Brown}, A.~G.~A. and {Prusti}, T. and {de Bruijne}, J.~H.~J. and {Arenou}, F. and {Babusiaux}, C. and {Biermann}, M. and {Creevey}, O.~L. and {Ducourant}, C. and {Evans}, D.~W. and {Eyer}, L. and {Guerra}, R. and {Hutton}, A. and {Jordi}, C. and {Klioner}, S.~A. and {Lammers}, U.~L. and {Lindegren}, L. and {Luri}, X. and {Mignard}, F. and {Panem}, C. and {Pourbaix}, D. and {Randich}, S. and {Sartoretti}, P. and {Soubiran}, C. and {Tanga}, P. and {Walton}, N.~A. and {Bailer-Jones}, C.~A.~L. and {Bastian}, U. and {Drimmel}, R. and {Jansen}, F. and {Katz}, D. and {Lattanzi}, M.~G. and {van Leeuwen}, F. and {Bakker}, J. and {Cacciari}, C. and {Casta{\~n}eda}, J. and {De Angeli}, F. and {Fabricius}, C. and {Fouesneau}, M. and {Fr{\'e}mat}, Y. and {Galluccio}, L. and {Guerrier}, A. and {Heiter}, U. and {Masana}, E. and {Messineo}, R. and {Mowlavi}, N. and {Nicolas}, C. and {Nienartowicz}, K. and {Pailler}, F. and {Panuzzo}, P. and {Riclet}, F. and {Roux}, W. and {Seabroke}, G.~M. and {Sordo}, R. and {Th{\'e}venin}, F. and {Gracia-Abril}, G. and {Portell}, J. and {Teyssier}, D. and {Altmann}, M. and {Andrae}, R. and {Audard}, M. and {Bellas-Velidis}, I. and {Benson}, K. and {Berthier}, J. and {Blomme}, R. and {Burgess}, P.~W. and {Busonero}, D. and {Busso}, G. and {C{\'a}novas}, H. and {Carry}, B. and {Cellino}, A. and {Cheek}, N. and {Clementini}, G. and {Damerdji}, Y. and {Davidson}, M. and {de Teodoro}, P. and {Nu{\~n}ez Campos}, M. and {Delchambre}, L. and {Dell'Oro}, A. and {Esquej}, P. and {Fern{\'a}ndez-Hern{\'a}ndez}, J. and {Fraile}, E. and {Garabato}, D. and {Garc{\'\i}a-Lario}, P. and {Gosset}, E. and {Haigron}, R. and {Halbwachs}, J.-L. and {Hambly}, N.~C. and {Harrison}, D.~L. and {Hern{\'a}ndez}, J. and {Hestroffer}, D. and {Hodgkin}, S.~T. and {Holl}, B. and {Jan{\ss}en}, K. and {Jevardat de Fombelle}, G. and {Jordan}, S. and {Krone-Martins}, A. and {Lanzafame}, A.~C. and {L{\"o}ffler}, W. and {Marchal}, O. and {Marrese}, P.~M. and {Moitinho}, A. and {Muinonen}, K. and {Osborne}, P. and {Pancino}, E. and {Pauwels}, T. and {Recio-Blanco}, A. and {Reyl{\'e}}, C. and {Riello}, M. and {Rimoldini}, L. and {Roegiers}, T. and {Rybizki}, J. and {Sarro}, L.~M. and {Siopis}, C. and {Smith}, M. and {Sozzetti}, A. and {Utrilla}, E. and {van Leeuwen}, M. and {Abbas}, U. and {{\'A}brah{\'a}m}, P. and {Abreu Aramburu}, A. and {Aerts}, C. and {Aguado}, J.~J. and {Ajaj}, M. and {Aldea-Montero}, F. and {Altavilla}, G. and {{\'A}lvarez}, M.~A. and {Alves}, J. and {Anders}, F. and {Anderson}, R.~I. and {Anglada Varela}, E. and {Antoja}, T. and {Baines}, D. and {Baker}, S.~G. and {Balaguer-N{\'u}{\~n}ez}, L. and {Balbinot}, E. and {Balog}, Z. and {Barache}, C. and {Barbato}, D. and {Barros}, M. and {Barstow}, M.~A. and {Bartolom{\'e}}, S. and {Bassilana}, J.-L. and {Bauchet}, N. and {Becciani}, U. and {Bellazzini}, M. and {Berihuete}, A. and {Bernet}, M. and {Bertone}, S. and {Bianchi}, L. and {Binnenfeld}, A. and {Blanco-Cuaresma}, S. and {Blazere}, A. and {Boch}, T. and {Bombrun}, A. and {Bossini}, D. and {Bouquillon}, S. and {Bragaglia}, A. and {Bramante}, L. and {Breedt}, E. and {Bressan}, A. and {Brouillet}, N. and {Brugaletta}, E. and {Bucciarelli}, B. and {Burlacu}, A. and {Butkevich}, A.~G. and {Buzzi}, R. and {Caffau}, E. and {Cancelliere}, R. and {Cantat-Gaudin}, T. and {Carballo}, R. and {Carlucci}, T. and {Carnerero}, M.~I. and {Carrasco}, J.~M. and {Casamiquela}, L. and {Castellani}, M. and {Castro-Ginard}, A. and {Chaoul}, L. and {Charlot}, P. and {Chemin}, L. and {Chiaramida}, V. and {Chiavassa}, A. and {Chornay}, N. and {Comoretto}, G. and {Contursi}, G. and {Cooper}, W.~J. and {Cornez}, T. and {Cowell}, S. and {Crifo}, F. and {Cropper}, M. and {Crosta}, M. and {Crowley}, C. and {Dafonte}, C. and {Dapergolas}, A. and {David}, M. and {David}, P. and {de Laverny}, P. and {De Luise}, F. and {De March}, R.},
        title = "{Gaia Data Release 3. Summary of the content and survey properties}",
      journal = {\aap},
         year = 2023,
        month = jun,
       volume = {674},
          eid = {A1},
        pages = {A1},
          doi = {10.1051/0004-6361/202243940},
archivePrefix = {arXiv},
       eprint = {2208.00211},
 primaryClass = {astro-ph.GA},
       adsurl = {https://ui.adsabs.harvard.edu/abs/2023A&A...674A...1G}
}

@ARTICLE{Drozdovskaya2016,
       author = {{Drozdovskaya}, Maria N. and {Walsh}, Catherine and {van Dishoeck}, Ewine F. and {Furuya}, Kenji and {Marboeuf}, Ulysse and {Thiabaud}, Amaury and {Harsono}, Daniel and {Visser}, Ruud},
        title = "{Cometary ices in forming protoplanetary disc midplanes}",
      journal = {\mnras},
         year = 2016,
        month = oct,
       volume = {462},
       number = {1},
        pages = {977-993},
          doi = {10.1093/mnras/stw1632},
archivePrefix = {arXiv},
       eprint = {1607.07861},
 primaryClass = {astro-ph.EP},
       adsurl = {https://ui.adsabs.harvard.edu/abs/2016MNRAS.462..977D}
}

@ARTICLE{Lichtenberg2021,
       author = {{Lichtenberg}, Tim and {Dr{\k{a}}{\.z}kowska}, Joanna and {Sch{\"o}nb{\"a}chler}, Maria and {Golabek}, Gregor J. and {Hands}, Thomas O.},
        title = "{Bifurcation of planetary building blocks during Solar System formation}",
      journal = {Science},
         year = 2021,
        month = jan,
       volume = {371},
       number = {6527},
        pages = {365-370},
          doi = {10.1126/science.abb3091},
archivePrefix = {arXiv},
       eprint = {2101.08571},
 primaryClass = {astro-ph.EP},
       adsurl = {https://ui.adsabs.harvard.edu/abs/2021Sci...371..365L}
}

@ARTICLE{Salyk2026,
       author = {{Salyk}, Colette and {Pontoppidan}, Klaus M. and {Zhang}, Ke and {Heinzen}, Sophie and {Calahan}, Jenny K. and {Banzatti}, Andrea and {Dickson-Vandervelde}, D. Annie and {Bergin}, Edwin A. and {Blake}, Geoffrey A. and {Arulanantham}, Nicole and {Krijt}, Sebastiaan and {Carr}, John and {Najita}, Joan and {Green}, Joel and {Romero-Mirza}, Carlos},
        title = "{H218O in the Terrestrial Planet-forming Regions of Protoplanetary Disks}",
      journal = {\apj},
         year = 2026,
        month = mar,
       volume = {999},
       number = {1},
          eid = {92},
        pages = {92},
          doi = {10.3847/1538-4357/ae3db1},
archivePrefix = {arXiv},
       eprint = {2601.20096},
 primaryClass = {astro-ph.EP},
       adsurl = {https://ui.adsabs.harvard.edu/abs/2026ApJ...999...92S}
}

@ARTICLE{Hayashi1981,
       author = {{Hayashi}, C.},
        title = "{Structure of the Solar Nebula, Growth and Decay of Magnetic Fields and Effects of Magnetic and Turbulent Viscosities on the Nebula}",
      journal = {Prog. Theor. Phys. Suppl.},
         year = 1981,
        month = jan,
       volume = {70},
        pages = {35-53},
          doi = {10.1143/PTPS.70.35},
       adsurl = {https://ui.adsabs.harvard.edu/abs/1981PThPS..70...35H}
}

@ARTICLE{Huet2025,
       author = {{Huet}, Paul and {Kral}, Quentin and {Guillot}, Tristan},
        title = "{Late gas released in the young Kuiper belt could have significantly contributed to the carbon enrichment of the atmospheres of Neptune and Uranus}",
      journal = {\aap},
         year = 2025,
        month = jul,
       volume = {699},
          eid = {A278},
        pages = {A278},
          doi = {10.1051/0004-6361/202554691},
archivePrefix = {arXiv},
       eprint = {2506.03412},
 primaryClass = {astro-ph.EP},
       adsurl = {https://ui.adsabs.harvard.edu/abs/2025A&A...699A.278H}
}

@ARTICLE{Notsu2018,
       author = {{Notsu}, Shota and {Nomura}, Hideko and {Walsh}, Catherine and {Honda}, Mitsuhiko and {Hirota}, Tomoya and {Akiyama}, Eiji and {Millar}, T.~J.},
        title = "{Candidate Water Vapor Lines to Locate the H$_{2}$O Snowline through High-dispersion Spectroscopic Observations. III. Submillimeter H$_{2}$ $^{16}$O and H$_{2}$ $^{18}$O Lines}",
      journal = {\apj},
         year = 2018,
        month = mar,
       volume = {855},
       number = {1},
          eid = {62},
        pages = {62},
          doi = {10.3847/1538-4357/aaaa72},
archivePrefix = {arXiv},
       eprint = {1801.07437},
 primaryClass = {astro-ph.EP},
       adsurl = {https://ui.adsabs.harvard.edu/abs/2018ApJ...855...62N}
}

@ARTICLE{Teague2019,
       author = {{Teague}, Richard},
        title = "{GoFish: Fishing for Line Observations in Protoplanetary Disks}",
      journal = {JOSS},
         year = 2019,
        month = sep,
       volume = {4},
       number = {41},
          eid = {1632},
        pages = {1632},
          doi = {10.21105/joss.01632},
       adsurl = {https://ui.adsabs.harvard.edu/abs/2019JOSS....4.1632T}
}

@ARTICLE{Yen2016,
       author = {{Yen}, Hsi-Wei and {Koch}, Patrick M. and {Liu}, Hauyu Baobab and {Puspitaningrum}, Evaria and {Hirano}, Naomi and {Lee}, Chin-Fei and {Takakuwa}, Shigehisa},
        title = "{Stacking Spectra in Protoplanetary Disks: Detecting Intensity Profiles from Hidden Molecular Lines in HD 163296}",
      journal = {\apj},
         year = 2016,
        month = dec,
       volume = {832},
       number = {2},
          eid = {204},
        pages = {204},
          doi = {10.3847/0004-637X/832/2/204},
archivePrefix = {arXiv},
       eprint = {1610.01780},
 primaryClass = {astro-ph.SR},
       adsurl = {https://ui.adsabs.harvard.edu/abs/2016ApJ...832..204Y}
}

@ARTICLE{Carr2018,
       author = {{Carr}, John S. and {Najita}, Joan R. and {Salyk}, Colette},
        title = "{Measuring the Water Snow Line in a Protoplanetary Disk}",
      journal = {RNAAS},
         year = 2018,
        month = sep,
       volume = {2},
       number = {3},
          eid = {169},
        pages = {169},
          doi = {10.3847/2515-5172/aadfe7},
       adsurl = {https://ui.adsabs.harvard.edu/abs/2018RNAAS...2..169C}
}

@ARTICLE{Bosman2021,
       author = {{Bosman}, Arthur D. and {Bergin}, Edwin A.},
        title = "{Reimagining the Water Snowline}",
      journal = {\apjl},
         year = 2021,
        month = sep,
       volume = {918},
       number = {1},
          eid = {L10},
        pages = {L10},
          doi = {10.3847/2041-8213/ac1db1},
archivePrefix = {arXiv},
       eprint = {2108.07303},
 primaryClass = {astro-ph.EP},
       adsurl = {https://ui.adsabs.harvard.edu/abs/2021ApJ...918L..10B}
}

@ARTICLE{Jones2011,
       author = {{Jones}, Brant M. and {Bennett}, Christopher J. and {Kaiser}, Ralf I.},
        title = "{Mechanistical Studies on the Production of Formamide (H$_{2}$NCHO) within Interstellar Ice Analogs}",
      journal = {\apj},
         year = 2011,
        month = jun,
       volume = {734},
       number = {2},
          eid = {78},
        pages = {78},
          doi = {10.1088/0004-637X/734/2/78},
       adsurl = {https://ui.adsabs.harvard.edu/abs/2011ApJ...734...78J}
}

@ARTICLE{Nomura2007,
       author = {{Nomura}, H. and {Aikawa}, Y. and {Tsujimoto}, M. and {Nakagawa}, Y. and {Millar}, T.~J.},
        title = "{Molecular Hydrogen Emission from Protoplanetary Disks. II. Effects of X-Ray Irradiation and Dust Evolution}",
      journal = {\apj},
         year = 2007,
        month = may,
       volume = {661},
       number = {1},
        pages = {334-353},
          doi = {10.1086/513419},
archivePrefix = {arXiv},
       eprint = {astro-ph/0702030},
 primaryClass = {astro-ph},
       adsurl = {https://ui.adsabs.harvard.edu/abs/2007ApJ...661..334N}
}

@ARTICLE{Favre2018,
       author = {{Favre}, C{\'e}cile and {Fedele}, Davide and {Semenov}, Dmitry and {Parfenov}, Sergey and {Codella}, Claudio and {Ceccarelli}, Cecilia and {Bergin}, Edwin A. and {Chapillon}, Edwige and {Testi}, Leonardo and {Hersant}, Franck and {Lefloch}, Bertrand and {Fontani}, Francesco and {Blake}, Geoffrey A. and {Cleeves}, L. Ilsedore and {Qi}, Chunhua and {Schwarz}, Kamber R. and {Taquet}, Vianney},
        title = "{First Detection of the Simplest Organic Acid in a Protoplanetary Disk}",
      journal = {\apjl},
         year = 2018,
        month = jul,
       volume = {862},
       number = {1},
          eid = {L2},
        pages = {L2},
          doi = {10.3847/2041-8213/aad046},
archivePrefix = {arXiv},
       eprint = {1807.05768},
 primaryClass = {astro-ph.SR},
       adsurl = {https://ui.adsabs.harvard.edu/abs/2018ApJ...862L...2F}
}

@ARTICLE{Jeong2025,
       author = {{Jeong}, Jae-Hong and {Lee}, Jeong-Eun and {Lee}, Seonjae and {Baek}, Giseon and {Kang}, Ji-Hyun and {Lee}, Seokho and {Kim}, Chul-Hwan and {Yun}, Hyeong-Sik and {Aikawa}, Yuri and {Herczeg}, Gregory J. and {Johnstone}, Doug and {Cieza}, Lucas},
        title = "{ALMA Spectral Survey of an Eruptive Young Star, V883 Ori (ASSAY). II. Freshly Sublimated Complex Organic Molecules in the Keplerian Disk}",
      journal = {\apjs},
         year = 2025,
        month = feb,
       volume = {276},
       number = {2},
          eid = {49},
        pages = {49},
          doi = {10.3847/1538-4365/ad9450},
archivePrefix = {arXiv},
       eprint = {2411.03826},
 primaryClass = {astro-ph.SR},
       adsurl = {https://ui.adsabs.harvard.edu/abs/2025ApJS..276...49J}
}

@ARTICLE{Yamato2024,
       author = {{Yamato}, Yoshihide and {Notsu}, Shota and {Aikawa}, Yuri and {Okoda}, Yuki and {Nomura}, Hideko and {Sakai}, Nami},
        title = "{Chemistry of Complex Organic Molecules in the V883 Ori Disk Revealed by ALMA Band 3 Observations}",
      journal = {\aj},
         year = 2024,
        month = feb,
       volume = {167},
       number = {2},
          eid = {66},
        pages = {66},
          doi = {10.3847/1538-3881/ad11d9},
archivePrefix = {arXiv},
       eprint = {2312.01300},
 primaryClass = {astro-ph.EP},
       adsurl = {https://ui.adsabs.harvard.edu/abs/2024AJ....167...66Y}
}

@ARTICLE{Muller2001,
       author = {{M{\"u}ller}, H.~S.~P. and {Thorwirth}, S. and {Roth}, D.~A. and {Winnewisser}, G.},
        title = "{The Cologne Database for Molecular Spectroscopy, CDMS}",
      journal = {\aap},
         year = 2001,
        month = apr,
       volume = {370},
        pages = {L49-L52},
          doi = {10.1051/0004-6361:20010367},
       adsurl = {https://ui.adsabs.harvard.edu/abs/2001A&A...370L..49M}
}

@ARTICLE{2023Natur.620..516P,
       author = {{Perotti}, G. and {Christiaens}, V. and {Henning}, Th. and {Tabone}, B. and {Waters}, L.~B.~F.~M. and {Kamp}, I. and {Olofsson}, G. and {Grant}, S.~L. and {Gasman}, D. and {Bouwman}, J. and {Samland}, M. and {Franceschi}, R. and {van Dishoeck}, E.~F. and {Schwarz}, K. and {G{\"u}del}, M. and {Lagage}, P.-O. and {Ray}, T.~P. and {Vandenbussche}, B. and {Abergel}, A. and {Absil}, O. and {Arabhavi}, A.~M. and {Argyriou}, I. and {Barrado}, D. and {Boccaletti}, A. and {Caratti o Garatti}, A. and {Geers}, V. and {Glauser}, A.~M. and {Justannont}, K. and {Lahuis}, F. and {Mueller}, M. and {Nehm{\'e}}, C. and {Pantin}, E. and {Scheithauer}, S. and {Waelkens}, C. and {Guadarrama}, R. and {Jang}, H. and {Kanwar}, J. and {Morales-Calder{\'o}n}, M. and {Pawellek}, N. and {Rodgers-Lee}, D. and {Schreiber}, J. and {Colina}, L. and {Greve}, T.~R. and {{\"O}stlin}, G. and {Wright}, G.},
        title = "{Water in the terrestrial planet-forming zone of the PDS 70 disk}",
      journal = {\nat},
         year = 2023,
        month = aug,
       volume = {620},
       number = {7974},
        pages = {516-520},
          doi = {10.1038/s41586-023-06317-9},
archivePrefix = {arXiv},
       eprint = {2307.12040},
 primaryClass = {astro-ph.EP},
       adsurl = {https://ui.adsabs.harvard.edu/abs/2023Natur.620..516P}
}

@ARTICLE{2026Natur.655..870C,
       author = {{Cordiner}, Martin and {Roth}, Nathan X. and {Micheli}, Marco and {Villanueva}, Geronimo and {Farnocchia}, Davide and {Charnley}, Steven and {Biver}, Nicolas and {Bockel{\'e}e-Morvan}, Dominique and {Bodewits}, Dennis and {Chandler}, Colin Orion and {Crovisier}, Jacques and {Drozdovskaya}, Maria N. and {Furuya}, Kenji and {Kelley}, Michael S.~P. and {Milam}, Stefanie and {Noonan}, John W. and {Opitom}, Cyrielle and {Schwamb}, Megan E. and {Thomas}, Cristina A.},
        title = "{Isotopic evidence for a cold and distant origin of 3I/ATLAS}",
      journal = {\nat},
         year = 2026,
        month = jul,
       volume = {655},
       number = {8124},
        pages = {870-874},
          doi = {10.1038/s41586-026-10771-6},
archivePrefix = {arXiv},
       eprint = {2603.06911},
 primaryClass = {astro-ph.EP},
       adsurl = {https://ui.adsabs.harvard.edu/abs/2026Natur.655..870C}
}

@ARTICLE{2025ARA&A..63...83S,
       author = {{Snellen}, Ignas A.~G.},
        title = "{Exoplanet Atmospheres at High Spectral Resolution}",
      journal = {\araa},
         year = 2025,
        month = aug,
       volume = {63},
       number = {1},
        pages = {83-125},
          doi = {10.1146/annurev-astro-052622-031342},
archivePrefix = {arXiv},
       eprint = {2505.08926},
 primaryClass = {astro-ph.EP},
       adsurl = {https://ui.adsabs.harvard.edu/abs/2025ARA&A..63...83S}
}

@ARTICLE{Walsh2014,
       author = {{Walsh}, Catherine and {Millar}, Tom. J. and {Nomura}, Hideko and {Herbst}, Eric and {Widicus Weaver}, Susanna and {Aikawa}, Yuri and {Laas}, Jacob C. and {Vasyunin}, Anton I.},
        title = "{Complex organic molecules in protoplanetary disks}",
      journal = {\aap},
         year = 2014,
        month = mar,
       volume = {563},
          eid = {A33},
        pages = {A33},
          doi = {10.1051/0004-6361/201322446},
archivePrefix = {arXiv},
       eprint = {1403.0390},
 primaryClass = {astro-ph.EP},
       adsurl = {https://ui.adsabs.harvard.edu/abs/2014A&A...563A..33W}
}

@INPROCEEDINGS{Walsh2018,
       author = {{Walsh}, Catherine and {Vissapragada}, Shreyas and {McGee}, Harry},
        title = "{Methanol formation in TW Hya and future prospects for detecting larger complex molecules in disks with ALMA}",
    booktitle = {Astrochemistry VII},
         year = 2018,
       editor = {{Cunningham}, Maria and et al.},
       series = {IAU Symposium},
       volume = {332},
        month = sep,
        pages = {395-402},
          doi = {10.1017/S1743921317007037},
archivePrefix = {arXiv},
       eprint = {1710.01219},
 primaryClass = {astro-ph.EP},
       adsurl = {https://ui.adsabs.harvard.edu/abs/2018IAUS..332..395W}
}

@ARTICLE{Booth2026,
       author = {{Booth}, Alice S. and {Calahan}, Jenny and {Temmink}, Milou and {W{\"o}lfer}, Lisa and {Pegues}, Jamila and {Law}, Charles J. and {Evans}, Lucy and {Leemker}, Margot and {Notsu}, Shota and {{\"O}berg}, Karin and {Walsh}, Catherine and {van Dishoeck}, Ewine F.},
        title = "{The Chemical Diversity of Giant-planet Nurseries as Revealed by ALMA}",
      journal = {\aj},
         year = 2026,
        month = mar,
       volume = {171},
       number = {3},
          eid = {128},
        pages = {128},
          doi = {10.3847/1538-3881/ae286b},
archivePrefix = {arXiv},
       eprint = {2512.01731},
 primaryClass = {astro-ph.EP},
       adsurl = {https://ui.adsabs.harvard.edu/abs/2026AJ....171..128B}
}

@ARTICLE{Lee2019,
       author = {{Lee}, Jeong-Eun and {Lee}, Seokho and {Baek}, Giseon and {Aikawa}, Yuri and {Cieza}, Lucas and {Yoon}, Sung-Yong and {Herczeg}, Gregory and {Johnstone}, Doug and {Casassus}, Simon},
        title = "{The ice composition in the disk around V883 Ori revealed by its stellar outburst}",
      journal = {Nat. Astron.},
         year = 2019,
        month = feb,
       volume = {3},
        pages = {314-319},
          doi = {10.1038/s41550-018-0680-0},
archivePrefix = {arXiv},
       eprint = {1809.00353},
 primaryClass = {astro-ph.SR},
       adsurl = {https://ui.adsabs.harvard.edu/abs/2019NatAs...3..314L}
}

@ARTICLE{Ilee2026,
       author = {{Ilee}, John D. and {Walsh}, Catherine and {Calahan}, Jenny K.},
        title = "{Methanol Emission Tracing Ice Chemistry and Dust Evolution in the TW Hya Protoplanetary Disk}",
      journal = {\aj},
         year = 2026,
        month = jan,
       volume = {171},
       number = {1},
          eid = {3},
        pages = {3},
          doi = {10.3847/1538-3881/ae100e},
archivePrefix = {arXiv},
       eprint = {2510.04106},
 primaryClass = {astro-ph.GA},
       adsurl = {https://ui.adsabs.harvard.edu/abs/2026AJ....171....3I}
}

@ARTICLE{Calahan2023,
       author = {{Calahan}, Jenny K. and {Bergin}, Edwin A. and {Bosman}, Arthur D. and {Rich}, Evan A. and {Andrews}, Sean M. and {Bergner}, Jennifer B. and {Cleeves}, L. Ilsedore and {Guzm{\'a}n}, Viviana V. and {Huang}, Jane and {Ilee}, John D. and {Law}, Charles J. and {Le Gal}, Romane and {{\"O}berg}, Karin I. and {Teague}, Richard and {Walsh}, Catherine and {Wilner}, David J. and {Zhang}, Ke},
        title = "{UV-driven chemistry as a signpost of late-stage planet formation}",
      journal = {Nat. Astron.},
         year = 2023,
        month = jan,
       volume = {7},
        pages = {49-56},
          doi = {10.1038/s41550-022-01831-8},
archivePrefix = {arXiv},
       eprint = {2212.05539},
 primaryClass = {astro-ph.EP},
       adsurl = {https://ui.adsabs.harvard.edu/abs/2023NatAs...7...49C}
}

@ARTICLE{Parfenov2017,
       author = {{Parfenov}, S. Yu. and {Semenov}, D.~A. and {Henning}, Th. and {Shapovalova}, A.~S. and {Sobolev}, A.~M. and {Teague}, R.},
        title = "{On the methanol emission detection in the TW Hya disc: the role of grain surface chemistry and non-LTE excitation}",
      journal = {\mnras},
         year = 2017,
        month = jun,
       volume = {468},
       number = {2},
        pages = {2024-2031},
          doi = {10.1093/mnras/stx624},
archivePrefix = {arXiv},
       eprint = {1703.03748},
 primaryClass = {astro-ph.SR},
       adsurl = {https://ui.adsabs.harvard.edu/abs/2017MNRAS.468.2024P}
}

@ARTICLE{Loomis2018,
       author = {{Loomis}, Ryan A. and {Cleeves}, L. Ilsedore and {{\"O}berg}, Karin I. and {Aikawa}, Yuri and {Bergner}, Jennifer and {Furuya}, Kenji and {Guzman}, V.~V. and {Walsh}, Catherine},
        title = "{The Distribution and Excitation of CH$_{3}$CN in a Solar Nebula Analog}",
      journal = {\apj},
         year = 2018,
        month = jun,
       volume = {859},
       number = {2},
          eid = {131},
        pages = {131},
          doi = {10.3847/1538-4357/aac169},
archivePrefix = {arXiv},
       eprint = {1805.01458},
 primaryClass = {astro-ph.EP},
       adsurl = {https://ui.adsabs.harvard.edu/abs/2018ApJ...859..131L}
}

@ARTICLE{Walsh2016,
       author = {{Walsh}, Catherine and {Loomis}, Ryan A. and {{\"O}berg}, Karin I. and {Kama}, Mihkel and {van 't Hoff}, Merel L.~R. and {Millar}, Tom J. and {Aikawa}, Yuri and {Herbst}, Eric and {Widicus Weaver}, Susanna L. and {Nomura}, Hideko},
        title = "{First Detection of Gas-phase Methanol in a Protoplanetary Disk}",
      journal = {\apjl},
         year = 2016,
        month = may,
       volume = {823},
       number = {1},
          eid = {L10},
        pages = {L10},
          doi = {10.3847/2041-8205/823/1/L10},
archivePrefix = {arXiv},
       eprint = {1606.06492},
 primaryClass = {astro-ph.EP},
       adsurl = {https://ui.adsabs.harvard.edu/abs/2016ApJ...823L..10W}
}

@ARTICLE{Schoier2005,
       author = {{Sch{\"o}ier}, F.~L. and {van der Tak}, F.~F.~S. and {van Dishoeck}, E.~F. and {Black}, J.~H.},
        title = "{An atomic and molecular database for analysis of submillimetre line observations}",
      journal = {\aap},
         year = 2005,
        month = mar,
       volume = {432},
       number = {1},
        pages = {369-379},
          doi = {10.1051/0004-6361:20041729},
archivePrefix = {arXiv},
       eprint = {astro-ph/0411110},
 primaryClass = {astro-ph},
       adsurl = {https://ui.adsabs.harvard.edu/abs/2005A&A...432..369S}
}

@ARTICLE{Cieza2016,
       author = {{Cieza}, Lucas A. and {Casassus}, Simon and {Tobin}, John and {Bos}, Steven P. and {Williams}, Jonathan P. and {Perez}, Sebastian and {Zhu}, Zhaohuan and {Caceres}, Claudio and {Canovas}, Hector and {Dunham}, Michael M. and {Hales}, Antonio and {Prieto}, Jose L. and {Principe}, David A. and {Schreiber}, Matthias R. and {Ruiz-Rodriguez}, Dary and {Zurlo}, Alice},
        title = "{Imaging the water snow-line during a protostellar outburst}",
      journal = {\nat},
         year = 2016,
        month = jul,
       volume = {535},
       number = {7611},
        pages = {258-261},
          doi = {10.1038/nature18612},
archivePrefix = {arXiv},
       eprint = {1607.03757},
 primaryClass = {astro-ph.SR},
       adsurl = {https://ui.adsabs.harvard.edu/abs/2016Natur.535..258C}
}

@ARTICLE{Booth2021,
       author = {{Booth}, Alice S. and {Walsh}, Catherine and {Terwisscha van Scheltinga}, Jeroen and {van Dishoeck}, Ewine F. and {Ilee}, John D. and {Hogerheijde}, Michiel R. and {Kama}, Mihkel and {Nomura}, Hideko},
        title = "{An inherited complex organic molecule reservoir in a warm planet-hosting disk}",
      journal = {Nat. Astron.},
         year = 2021,
        month = jan,
       volume = {5},
        pages = {684-690},
          doi = {10.1038/s41550-021-01352-w},
archivePrefix = {arXiv},
       eprint = {2104.08348},
 primaryClass = {astro-ph.EP},
       adsurl = {https://ui.adsabs.harvard.edu/abs/2021NatAs...5..684B}
}

@ARTICLE{Rampinelli2026,
       author = {{Rampinelli}, Luna and {Facchini}, Stefano and {Leemker}, Margot and {Isella}, Andrea and {Curone}, Pietro and {Benisty}, Myriam and {Humphreys}, Elizabeth and {Testi}, Leonardo},
        title = "{Water Vapor Emission at the Warm Cavity Wall of the HD 100546 Disk as Revealed by ALMA}",
      journal = {\apjl},
         year = 2026,
        month = jan,
       volume = {996},
       number = {1},
          eid = {L17},
        pages = {L17},
          doi = {10.3847/2041-8213/ae2868},
archivePrefix = {arXiv},
       eprint = {2512.06439},
 primaryClass = {astro-ph.EP},
       adsurl = {https://ui.adsabs.harvard.edu/abs/2026ApJ...996L..17R}
}

@ARTICLE{Leemker2025,
       author = {{Leemker}, Margot and {Tobin}, John J. and {Facchini}, Stefano and {Curone}, Pietro and {Booth}, Alice S. and {Furuya}, Kenji and {van't Hoff}, Merel L.~R.},
        title = "{Pristine ices in a planet-forming disk revealed by heavy water}",
      journal = {Nat. Astron.},
         year = 2025,
        month = oct,
       volume = {9},
        pages = {1486-1494},
          doi = {10.1038/s41550-025-02663-y},
archivePrefix = {arXiv},
       eprint = {2510.19919},
 primaryClass = {astro-ph.EP},
       adsurl = {https://ui.adsabs.harvard.edu/abs/2025NatAs...9.1486L}
}

@ARTICLE{Tobin2023,
       author = {{Tobin}, John J. and {van't Hoff}, Merel L.~R. and {Leemker}, Margot and {van Dishoeck}, Ewine F. and {Paneque-Carre{\~n}o}, Teresa and {Furuya}, Kenji and {Harsono}, Daniel and {Persson}, Magnus V. and {Cleeves}, L. Ilsedore and {Sheehan}, Patrick D. and {Cieza}, Lucas},
        title = "{Deuterium-enriched water ties planet-forming disks to comets and protostars}",
      journal = {\nat},
         year = 2023,
        month = mar,
       volume = {615},
       number = {7951},
        pages = {227-230},
          doi = {10.1038/s41586-022-05676-z},
       adsurl = {https://ui.adsabs.harvard.edu/abs/2023Natur.615..227T}
}

@ARTICLE{Leemker2026,
       author = {{Leemker}, M. and {Facchini}, S. and {Curone}, P. and {Rampinelli}, L. and {Benisty}, M. and {Garufi}, A. and {Humphreys}, E.},
        title = "{Zooming into the water snow line: High-resolution water observations of the HL Tau disk}",
      journal = {\aap},
         year = 2026,
        month = jan,
       volume = {705},
          eid = {A193},
        pages = {A193},
          doi = {10.1051/0004-6361/202557609},
archivePrefix = {arXiv},
       eprint = {2511.16737},
 primaryClass = {astro-ph.EP},
       adsurl = {https://ui.adsabs.harvard.edu/abs/2026A&A...705A.193L}
}

@ARTICLE{Facchini2024,
       author = {{Facchini}, Stefano and {Testi}, Leonardo and {Humphreys}, Elizabeth and {Vander Donckt}, Mathieu and {Isella}, Andrea and {Wrzosek}, Ramon and {Baudry}, Alain and {Gray}, Malcom D. and {Richards}, Anita M.~S. and {Vlemmings}, Wouter},
        title = "{Resolved ALMA observations of water in the inner astronomical units of the HL Tau disk}",
      journal = {Nat. Astron.},
         year = 2024,
        month = may,
       volume = {8},
        pages = {587-595},
          doi = {10.1038/s41550-024-02207-w},
archivePrefix = {arXiv},
       eprint = {2403.00647},
 primaryClass = {astro-ph.EP},
       adsurl = {https://ui.adsabs.harvard.edu/abs/2024NatAs...8..587F}
}

@ARTICLE{Bruderer2013,
       author = {{Bruderer}, Simon},
        title = "{Survival of molecular gas in cavities of transition disks. I. CO}",
      journal = {\aap},
         year = 2013,
        month = nov,
       volume = {559},
          eid = {A46},
        pages = {A46},
          doi = {10.1051/0004-6361/201321171},
archivePrefix = {arXiv},
       eprint = {1308.2966},
 primaryClass = {astro-ph.SR},
       adsurl = {https://ui.adsabs.harvard.edu/abs/2013A&A...559A..46B}
}

@ARTICLE{Bruderer2012,
       author = {{Bruderer}, S. and {van Dishoeck}, E.~F. and {Doty}, S.~D. and {Herczeg}, G.~J.},
        title = "{The warm gas atmosphere of the HD 100546 disk seen by Herschel. Evidence of a gas-rich, carbon-poor atmosphere?}",
      journal = {\aap},
         year = 2012,
        month = may,
       volume = {541},
          eid = {A91},
        pages = {A91},
          doi = {10.1051/0004-6361/201118218},
archivePrefix = {arXiv},
       eprint = {1201.4860},
 primaryClass = {astro-ph.SR},
       adsurl = {https://ui.adsabs.harvard.edu/abs/2012A&A...541A..91B}
}

@ARTICLE{Oberg2021b,
       author = {{\"Oberg}, Karin I. and {Bergin}, Edwin A.},
        title = "{Astrochemistry and compositions of planetary systems}",
      journal = {\physrep},
         year = 2021,
        month = jan,
       volume = {893},
        pages = {1-48},
          doi = {10.1016/j.physrep.2020.09.004},
archivePrefix = {arXiv},
       eprint = {2010.03529},
 primaryClass = {astro-ph.EP},
       adsurl = {https://ui.adsabs.harvard.edu/abs/2021PhR...893....1O}
}

@INPROCEEDINGS{Nomura2023,
       author = {{Nomura}, H. and {Furuya}, K. and {Cordiner}, M.~A. and {Charnley}, S.~B. and {Alexander}, C.~M. O'D. and {Nixon}, C.~A. and {Guzman}, V.~V. and {Yurimoto}, H. and {Tsukagoshi}, T. and {Iino}, T.},
        title = "{The Isotopic Links from Planet Forming Regions to the Solar System}",
    booktitle = {PPVII},
         year = 2023,
       volume = {534},
        month = jul,
        pages = {1075},
       adsurl = {https://ui.adsabs.harvard.edu/abs/2023ASPC..534.1075N}
}

@ARTICLE{Fortney10,
       author = {{Fortney}, Jonathan J. and {Nettelmann}, Nadine},
        title = "{The Interior Structure, Composition, and Evolution of Giant Planets}",
      journal = {\ssr},
         year = 2010,
        month = may,
       volume = {152},
       number = {1-4},
        pages = {423-447},
          doi = {10.1007/s11214-009-9582-x},
archivePrefix = {arXiv},
       eprint = {0912.0533},
 primaryClass = {astro-ph.EP},
       adsurl = {https://ui.adsabs.harvard.edu/abs/2010SSRv..152..423F}
}

@ARTICLE{Altwegg2019,
       author = {{Altwegg}, Kathrin and {Balsiger}, Hans and {Fuselier}, Stephen A.},
        title = "{Cometary Chemistry and the Origin of Icy Solar System Bodies: The View After Rosetta}",
      journal = {\araa},
         year = 2019,
        month = aug,
       volume = {57},
        pages = {113-155},
          doi = {10.1146/annurev-astro-091918-104409},
archivePrefix = {arXiv},
       eprint = {1908.04046},
 primaryClass = {astro-ph.EP},
       adsurl = {https://ui.adsabs.harvard.edu/abs/2019ARA&A..57..113A}
}

@ARTICLE{Nomura2021,
       author = {{Nomura}, Hideko and {Tsukagoshi}, Takashi and {Kawabe}, Ryohei and {Muto}, Takayuki and {Kanagawa}, Kazuhiro D. and {Aikawa}, Yuri and {Akiyama}, Eiji and {Okuzumi}, Satoshi and {Ida}, Shigeru and {Lee}, Seokho and {Walsh}, Catherine and {Millar}, T.~J.},
        title = "{High Spatial Resolution Observations of Molecular Lines toward the Protoplanetary Disk around TW Hya with ALMA}",
      journal = {\apj},
         year = 2021,
        month = jun,
       volume = {914},
       number = {2},
          eid = {113},
        pages = {113},
          doi = {10.3847/1538-4357/abfb6a},
archivePrefix = {arXiv},
       eprint = {2105.00976},
 primaryClass = {astro-ph.EP},
       adsurl = {https://ui.adsabs.harvard.edu/abs/2021ApJ...914..113N}
}

@ARTICLE{Garufi2021,
       author = {{Garufi}, A. and {Podio}, L. and {Codella}, C. and {Fedele}, D. and {Bianchi}, E. and {Favre}, C. and {Bacciotti}, F. and {Ceccarelli}, C. and {Mercimek}, S. and {Rygl}, K. and {Teague}, R. and {Testi}, L.},
        title = "{ALMA chemical survey of disk-outflow sources in Taurus (ALMA-DOT). V. Sample, overview, and demography of disk molecular emission}",
      journal = {\aap},
         year = 2021,
        month = jan,
       volume = {645},
          eid = {A145},
        pages = {A145},
          doi = {10.1051/0004-6361/202039483},
archivePrefix = {arXiv},
       eprint = {2012.07667},
 primaryClass = {astro-ph.GA},
       adsurl = {https://ui.adsabs.harvard.edu/abs/2021A&A...645A.145G}
}

@ARTICLE{Banzatti2025,
       author = {{Banzatti}, Andrea and {Salyk}, Colette and {Pontoppidan}, Klaus M. and {Carr}, John S. and {Zhang}, Ke and {Arulanantham}, Nicole and {Krijt}, Sebastiaan and {{\"O}berg}, Karin I. and {Cleeves}, L. Ilsedore and {Najita}, Joan R. and {Pascucci}, Ilaria and {Blake}, Geoffrey A. and {Romero-Mirza}, Carlos E. and {Bergin}, Edwin A. and {Cieza}, Lucas A. and {Pinilla}, Paola and {Long}, Feng and {Mallaney}, Patrick and {Xie}, Chengyan and {Waggoner}, Abygail R. and {Kaeufer}, Till and {The Jdiscs Collaboration}},
        title = "{Water in Protoplanetary Disks with JWST-MIRI: Spectral Excitation Atlas and Radial Distribution from Temperature Diagnostic Diagrams and Doppler Mapping}",
      journal = {\aj},
         year = 2025,
        month = mar,
       volume = {169},
       number = {3},
          eid = {165},
        pages = {165},
          doi = {10.3847/1538-3881/ada962},
archivePrefix = {arXiv},
       eprint = {2409.16255},
 primaryClass = {astro-ph.EP},
       adsurl = {https://ui.adsabs.harvard.edu/abs/2025AJ....169..165B}
}

@ARTICLE{Tabone2023,
       author = {{Tabone}, B. and {Bettoni}, G. and {van Dishoeck}, E.~F. and {Arabhavi}, A.~M. and {Grant}, S. and {Gasman}, D. and {Henning}, Th. and {Kamp}, I. and {G{\"u}del}, M. and {Lagage}, P.~O. and {Ray}, T. and {Vandenbussche}, B. and {Abergel}, A. and {Absil}, O. and {Argyriou}, I. and {Barrado}, D. and {Boccaletti}, A. and {Bouwman}, J. and {Caratti o Garatti}, A. and {Geers}, V. and {Glauser}, A.~M. and {Justannont}, K. and {Lahuis}, F. and {Mueller}, M. and {Nehm{\'e}}, C. and {Olofsson}, G. and {Pantin}, E. and {Scheithauer}, S. and {Waelkens}, C. and {Waters}, L.~B.~F.~M. and {Black}, J.~H. and {Christiaens}, V. and {Guadarrama}, R. and {Morales-Calder{\'o}n}, M. and {Jang}, H. and {Kanwar}, J. and {Pawellek}, N. and {Perotti}, G. and {Perrin}, A. and {Rodgers-Lee}, D. and {Samland}, M. and {Schreiber}, J. and {Schwarz}, K. and {Colina}, L. and {{\"O}stlin}, G. and {Wright}, G.},
        title = "{A rich hydrocarbon chemistry and high C to O ratio in the inner disk around a very low-mass star}",
      journal = {Nat. Astron.},
         year = 2023,
        month = jul,
       volume = {7},
        pages = {805-814},
          doi = {10.1038/s41550-023-01965-3},
archivePrefix = {arXiv},
       eprint = {2304.05954},
 primaryClass = {astro-ph.EP},
       adsurl = {https://ui.adsabs.harvard.edu/abs/2023NatAs...7..805T}
}

@ARTICLE{Alexander2018,
       author = {{Alexander}, Conel M. O'D. and {McKeegan}, Kevin D. and {Altwegg}, Kathrin},
        title = "{Water Reservoirs in Small Planetary Bodies: Meteorites, Asteroids, and Comets}",
      journal = {\ssr},
         year = 2018,
        month = feb,
       volume = {214},
       number = {1},
          eid = {36},
        pages = {36},
          doi = {10.1007/s11214-018-0474-9},
       adsurl = {https://ui.adsabs.harvard.edu/abs/2018SSRv..214...36A}
}
\end{document}